\documentclass[conference]{IEEEtran}
\IEEEoverridecommandlockouts
\usepackage{cite}
\usepackage{amsmath,amssymb,amsfonts}
\usepackage{algorithmic}
\usepackage{graphicx}
\usepackage{textcomp}
\usepackage{xcolor}
\usepackage{amsmath}
\usepackage{comment}
\usepackage[linesnumbered,ruled,vlined]{algorithm2e}
\usepackage{subcaption}
\usepackage{enumitem}
\usepackage{hyperref}
\usepackage{graphicx}
\usepackage{tabularx}
\usepackage{array}
\usepackage{multirow}
\usepackage{tabularx}

\newcolumntype{C}[1]{>{\centering\arraybackslash}p{#1}}
\newcolumntype{M}[1]{>{\centering\arraybackslash}m{#1}}

\def\BibTeX{{\rm B\kern-.05em{\sc i\kern-.025em b}\kern-.08em
    T\kern-.1667em\lower.7ex\hbox{E}\kern-.125emX}}
\begin{document}

\title{Transforming Agriculture with Intelligent Data Management and Insights
}

\author{

\IEEEauthorblockN{Yu Pan$^1$, Jianxin Sun$^2$, Hongfeng Yu$^{2,4}$, Geng Bai$^1$, Yufeng Ge$^1$, Joe Luck$^1$, Tala Awada$^3$}
\IEEEauthorblockA{$^1$\textit{Department of Biological System Engineering, University of Nebraska-Lincoln, Lincoln, NE, USA} \\
$^2$\textit{School of Computing, University of Nebraska-Lincoln, Lincoln, NE, USA}\\
$^3$\textit{School of Natural Resources, University of Nebraska-Lincoln, Lincoln, NE, USA}\\
$^4$\textit{Holland Computing Center, University of Nebraska-Lincoln, Lincoln, NE, USA}
}
}

\maketitle

\begin{abstract}
Modern agriculture faces grand challenges to meet increased demands for food, fuel, feed, and fiber with population growth under the constraints of climate change and dwindling natural resources. Data innovation is urgently required to secure and improve the productivity, sustainability, and resilience of our agroecosystems. As various sensors and Internet of Things (IoT) instrumentation become more available, affordable, reliable, and stable, it has become possible to conduct data collection, integration, and analysis at multiple temporal and spatial scales, in real-time, and with high resolutions. At the same time, the sheer amount of data poses a great challenge to data storage and analysis, and the \textit{de facto} data management and analysis practices adopted by scientists have become increasingly inefficient. Additionally, the data generated from different disciplines, such as genomics, phenomics, environment, agronomy, and socioeconomic, can be highly heterogeneous. That is, datasets across disciplines often do not share the same ontology, modality, or format. All of the above make it necessary to design a new data management infrastructure that implements the principles of Findable, Accessible, Interoperable, and Reusable (FAIR). In this paper, we propose Agriculture Data Management and Analytics (ADMA), which satisfies the FAIR principles. Our new data management infrastructure is intelligent by supporting semantic data management across disciplines, interactive by providing various data management/analysis portals such as web GUI, command line, and API, scalable by utilizing the power of high-performance computing (HPC), extensible by allowing users to load their own data analysis tools, trackable by keeping track of different operations on each file, and open by using a rich set of mature open source technologies.
\end{abstract}


\section{Introduction}

Modern agriculture stands at the crossroads of burgeoning global demands and a series of unprecedented challenges. As the world's population continues to grow, the agricultural sector is under increasing pressure to meet the escalating needs for food, fuel, feed, and fiber. This task is further complicated by the constraints of climate change, diminishing natural resources, and the intricate dynamics of global ecosystems. In this complex landscape, the role of data-informed decision-making becomes paramount. On the other hand, the proliferation of sensors, Internet of Things (IoT) devices, and advanced instrumentation has led to an explosion of data at various temporal and spatial scales. This vast amount of data holds the potential to revolutionize agricultural practices, offering insights into crop yields, soil health, and sustainable farming practices. 

However, the sheer volume and heterogeneity of the data present significant challenges in storage, integration, and analysis. Here we list the three main challenges faced by modern agriculture data management:

\begin{itemize}[leftmargin=*]

\item Massive volume. An unprecedented amount of datasets with multiple spatiotemporal resolutions are being collected from various sensors and IoT instrumentation.

\item Data heterogeneity. The data generated from different disciplines, such as genomics, phenomics, environment, agronomy, and socioeconomic, can be highly heterogeneous. Also, even for data from the same discipline, normally, they may have different formats and resolutions without spatiotemporal alignment.

\item In-transit processing. Current data processing practice usually involves several times of transmission of data. The data is collected in one place and then transmitted to another place for further processing, and the final results may be transmitted to a third place to visualize, which we call in-transit processing. The fundamental problem is that storage, processing, and visualization can be carried out in different cyberinfrastructure systems, which is increasingly inefficient when the data grows huge.
\end{itemize}

Efficient and intelligent data management and analytic infrastructures are now indispensable for harnessing the full potential of this information, driving the next wave of innovations in agriculture, and ensuring food security for future generations. There are existing agriculture data management projects being proposed and implemented \cite{swetnam2023cyverse,senay2022big,jouini2021evaluation,lebauer2020terra}, which answer specific requirements of agriculture research. However, due to the fast pace of technological advancement, the incorporation of new technologies, such as semantic \& conversational data search, is barely seen. The lack of support semantic search does not only influence the interactivity of these platforms but also has a fundamental impact on their implementation of FAIR principle.

To meet the challenges proposed by modern agriculture data management, we design and implement Agriculture Data Management and Analytics (ADMA). ADMA is a comprehensive platform developed specifically for experts in agriculture-related domains, which harnesses the power of advanced technologies, such as IoT, high-resolution sensors, and cloud computing, to facilitate data collection, integration, and analysis across various agricultural disciplines. It is a state-of-the-art solution designed to address the multifaceted challenges faced by modern agriculture. By incorporating the FAIR principles (Findable, Accessible, Interoperable, and Reusable), ADMA aims to break down data silos and promote collaboration among researchers, policymakers, and industry stakeholders. Key features of the ADMA platform include:

\begin{itemize}[leftmargin=*]

\item Intelligence: by incorporating Natural Language Processing (NLP)-generated high dimensional representations and incorporating vector data stores, ADMA supports semantic search across data formats and modalities.
\item Interactivity: accessible through web GUI, command line, and API, these user-friendly interfaces empower users to manage, analyze, and share data with ease.
\item Scalability: leveraging the HPC capabilities, ADMA can efficiently process large volumes of heterogeneous data.
\item Extensibility: ADMA allows users to incorporate their own data management and analysis tools. Also, users can plug in external data resources to ADMA and manage the external data and native data as if they are stored in the same place.
\item Open-source: built upon a robust framework of open-source technologies, ADMA fosters a collaborative environment for continuous improvement and innovation.
\item Trackability: ADMA keeps track of different operations on each file and maintains and visualizes the pipeline record for each file.
\item Privacy and Security: ADMA protects data privacy and guarantees data security by a variety of measures, such as separating data and metadata, differentiating public and private data, using containerized environments, and a unified authentication module.
\end{itemize}

By providing an integrated platform for agricultural data management and analysis, ADMA is poised to play a pivotal role in enhancing the productivity, sustainability, and resilience of our agroecosystems. Join us in revolutionizing the future of agriculture through data-driven insights and informed decision-making.

The contributions of our work are summarized as follows:
\begin{itemize}[leftmargin=*]

\item We design and implement Agriculture Data Management and Analytics (ADMA), which is an intelligent data management platform for all disciplines in agriculture research and education. 
  
 \item ADMA follows the FAIR principles and provides key features such as intelligence, interactivity, scalability and flexibility, open-source, pipeline management, and privacy and security.

\item Extensive demos and evaluations for ADMA are conducted, in which comparisons with existing data management platforms are made to show the innovation of our system.
\end{itemize}

The rest of the paper is organized as follows: Section \ref{sec:related} surveys existing work relevant to our research. Section \ref{sec:design} presents the design principles and high-level frameworks of our system. Section \ref{sec:implementation} introduces the implementation details of ADMA. In Section \ref{sec:evaluation}, we provide the thorough demos and evaluations of the current status of ADMA. Finally, in Section \ref{sec:conclusion}, we conclude the paper by discussing the contributions and suggesting future directions.
\section{Related Work}
\label{sec:related}

\subsection{Data Management System for Agricultural Research}

The digital transformation of agriculture has led to the development of sophisticated data management platforms tailored to address the unique challenges and requirements of agricultural research and production. There are several prominent platforms in related domains.

CyVerse \cite{merchant2016iplant},  originally developed for plant genomics, has evolved to provide life scientists with powerful computational infrastructure to handle vast datasets and intricate analyses. By offering a cloud-based platform, CyVerse facilitates collaborative research, enabling scientists to share data and tools seamlessly. The platform's emphasis on scalability and interoperability makes it a cornerstone in modern agricultural data management. GARDIAN \cite{jouini2021evaluation}, serving as the CGIAR (Consultative Group for International Agricultural Research)'s data discovery platform, enables the discovery of research datasets and publications across the CGIAR’s Centers and Programs. By integrating datasets from diverse agricultural research initiatives, GARDIAN provides a comprehensive view of global agricultural research data, promoting interdisciplinary research and collaboration. GEMS \cite{senay2022big} stands out for integrating genomic, phenotypic, and environmental data. By providing tools for the analysis and visualization of large-scale agricultural data, GEMS facilitates comprehensive studies, bridging the gap between genomics and environmental factors in agricultural research. TERRAREF \cite{lebauer2020terra} is a pioneering effort to provide high-resolution sensor data for plant research. Utilizing a plethora of sensors, including cameras and drones, TERRA-REF captures detailed data on plant growth, health, and environment. This rich dataset is a valuable resource for researchers aiming to understand the intricate interplay between plants and their environment. From Smart Farming towards Agriculture 5.0 \cite{saiz2020smart} discussed the evolution of Smart Farming and highlighted the importance of data in modern agriculture. It reviewed the current status of advanced farm management systems, emphasizing the role of data in optimizing decisions to ensure sustainability and economic efficiency.

There are also platforms and studies underscoring the significance of integrated data management systems in specific topics in agriculture or by using specific technologies. Agricultural Remote Sensing Big Data \cite{huang2018agricultural} discussed the management and applications of agricultural remote sensing big data. It introduced a four-layer-twelve-level (FLTL) remote sensing data management structure for managing and applying agricultural remote sensing big data for precision agriculture and local farm studies. Data warehouse and decision support on integrated crop big data 
\cite{ngo2020data} provided a comprehensive insight into the potential of data warehousing and decision support systems in harnessing the power of integrated crop big data by implementing a continental-level agricultural data warehouse (ADW). Big Data and Machine Learning (ML) With Hyperspectral Information in Agriculture \cite{ang2021big} reviewed significant research efforts in agriculture using Big data, machine learning, and deep learning, focusing on hyperspectral and multispectral data processing. It emphasized the potential of ensemble machine learning and scalable parallel discriminant analysis for Big data in agriculture. Blockchain for sustainable e-agriculture \cite{dey2021blockchain} discussed the integration of blockchain and IoT in e-agriculture. It highlighted the potential of this integration to improve linkages in agricultural value chains, benefit value chain actors, and enhance the performance of the IoT network. The study emphasized the role of blockchain in ensuring data validation, storage, security, and privacy in e-agriculture. 

\subsection{Semantic Search}

Semantic search has been a topic of interest for many years, aiming to improve the accuracy of information retrieval by understanding the intent and contextual meaning of user queries. Unlike traditional Information Retrieval (IR) Systems \cite{brin1998anatomy}, which rely on keyword matching and return results based on the frequency of terms, Semantic information retrieval systems avoid missing relevant documents that use different terminology by considering the semantic meaning of the content. There are mainly two types of methods approaching the problem of semantic search, namely, graph-based methods and embedding-based methods.

The use of ontologies and knowledge graphs \cite{chen2020review,chen2020knowledge} has been proposed to enhance search capabilities. These structures capture semantic relationships between entities, allowing for more contextually relevant search results. Notably, Google's Knowledge Graph is an example of this approach in action.

The other category, which gradually becomes the mainstream solution, is embedding-based methods. With the advent of deep learning, word embeddings like Word2Vec \cite{mikolov2013efficient}, GloVe~\cite{pennington2014glove}, and FastText \cite{joulin2016fasttext} have been developed to capture semantic relationships between words in a vector space. These embeddings have been employed in semantic search systems to measure the similarity between query and document embeddings. As labeled data is often scarce, zero-shot and few-shot learning techniques \cite{xian2018zero} have been explored for semantic search. These methods allow models to generalize and retrieve relevant information without extensive training data. More recently, transformer architectures, especially BERT (Bidirectional Encoder Representations from Transformers) \cite{devlin2018bert} and its variants, have set new benchmarks in understanding the context of words in a sentence. These models can be fine-tuned for semantic search tasks, offering significant improvements over traditional methods. With the proliferation of multimedia content, semantic search has also expanded to include images, videos, and audio. Multimodal search systems \cite{baltruvsaitis2018multimodal,tautkute2019deepstyle} aim to understand the content across different modalities and retrieve relevant information based on semantic understanding.

Despite the advancements, challenges remain in semantic search, including handling ambiguous queries, ensuring real-time search performance, and understanding cross-lingual semantics. Recent research has also delved into integrating common-sense reasoning and external knowledge bases to enhance semantic search capabilities further~\cite{lv2020graph}.

Like any data management system, information retrieval in agriculture data management systems plays a central role. By adopting semantic search empowered by large language models, users of the agriculture data management system can find relevant data and tools more accurately and efficiently. 

\section{Design}
\label{sec:design}

\subsection{Concepts}
\label{subsec: concepts}

ADMA maintains four modes of entities, namely data, tool, model, and collection, as the first-class species in the system.

\begin{itemize}[leftmargin=*] 
\item Data is the most general and default mode.
\item Tool represents both built-in and user-provided programs or scripts that process data and can be assembled into diverse pipelines.
\item Model represents files resulting from a machine learning training process, blending data and tools. During training with a training program, it acts as data; during inference, it serves as a tool.
\item Collection is a set of data, tools, and models, which users can customize by adding their own data, tools, and models.
\end{itemize}

\subsection{Framework}

\begin{figure}[t]
\centerline{\includegraphics[width=1.0\linewidth]{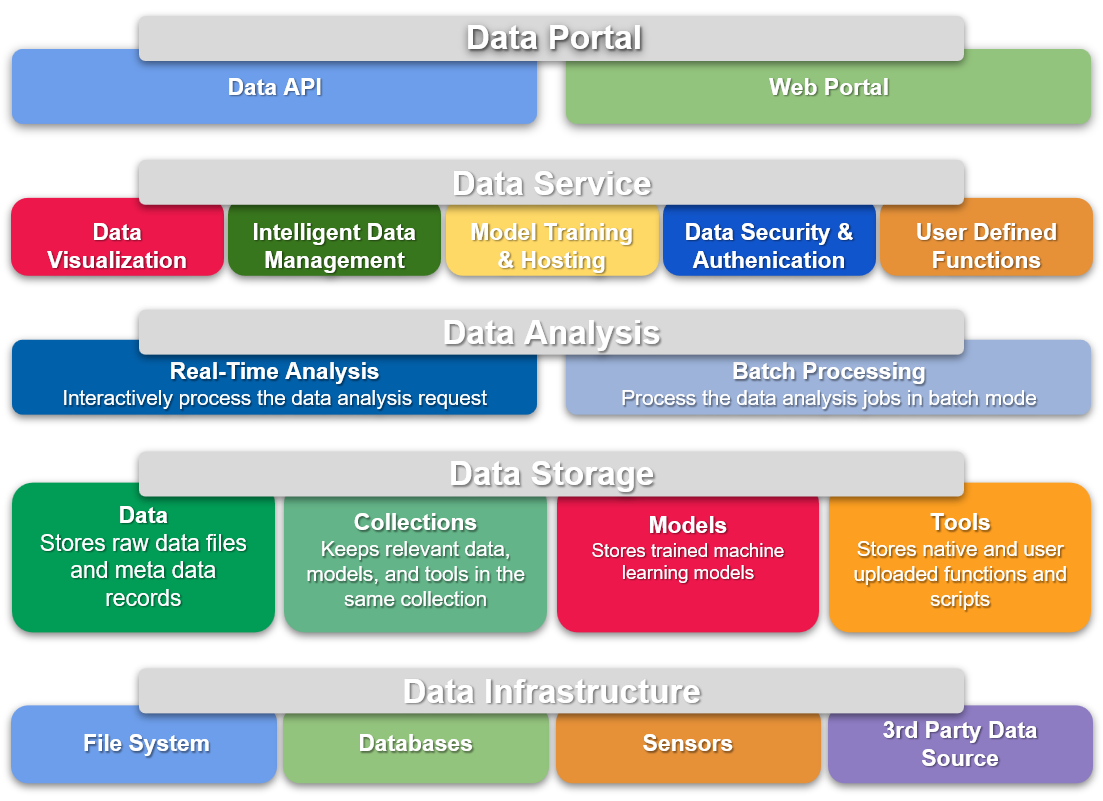}}
\caption{ADMA Framework}
\label{fig:framework}
\end{figure}

To support the FAIR principles and the proposed key features such as intelligence, interactivity, scalability and flexibility, open-source, and pipeline management, we design our ADAM in a top-down manner, and logically, there are five layers in the framework, as shown in Figure~\ref{fig:framework}. 

\subsubsection{Data Portal}

The data portal is the interface of the whole system, through which end users and 3rd party programs interact with the system and manipulate the data stored in the system. Currently, we provide two types of data portals: data API and web portal. 

The former exposes some of the basic data manipulation functionalities, such as searching, retrieving, uploading, and deleting the data and associated metadata. External systems or apps can programmatically make calls to the data APIs. One of the possible application scenarios is a 3rd party system that is responsible for collecting the data and can call APIs to upload the new data once it becomes available. Another application scenario is that an application can call APIs to pull the data from ADMA and visualize the data on the user's screen. 

The latter provides a user-friendly web GUI for end users to manipulate the data. The design ideology is that both portals should support the complete set of basic data management functionalities rather than complement each other. That is, if a task can be completed by using the web portal, it should also be completed by making calls to data API.

\subsubsection{Data Service}
\label{subsubsec: data_service}

The data service layer provides various native data manipulation modules such as data visualization, intelligent data management, model training and hosting, and data security and authentication. This layer also supports user-defined functions to provide some degree of extensibility of the whole system. Essentially, the data service layer is where most of the logic of the system resides. 

ADMA supports basic visualization or rendering of various data formats such as .shp, .tif, .xlsx, .csv, .py, .r, and so on. The intelligent data management module keeps all the data files of various formats and modalities in a unified format by mapping the metadata and data itself to the same high dimensional space by using NLP and other ML models. Then, these high dimensional vectors are stored in a vector data store, which supports a semantic search of the underlying data. In this way, users can search the data in various formats and modalities by using natural language without worrying that their query may not exactly match the data they want to retrieve. By keeping a representation of each data file in the vector data store, ADMA supports semantic data management instead of label-based, metadata-based, or path-based data management. As machine learning methodologies become ubiquitous in agricultural research, it is necessary to support model training and hosting services in ADMA. This module allows the users to natively train their ML models by feeding in datasets stored in ADMA and then storing, running, and hosting trained models on ADMA. ADMA supports multiple users and allows them to share their data or keep their data private by providing data security and security modules. This holds great significance as cross-disciplinary and inter-organizational collaboration increases, and the necessity to distinguish between shareable and proprietary data becomes pressing. User-defined functions or tools are also allowed to run on ADMA. As discussed in Section \ref{sec:implementation}, for safety, all the user-defined functions will be run in a separate container.

\subsubsection{Data Analysis}

Most of the modules in the data service operate within this layer, which primarily includes two types of data analysis: real-time analysis and batch processing. The former interactively processes the data analysis request and emphasizes the latency of each interactive task. A typical application scenario of real-time analysis is that ADMA continuously pulls data from some external data source (e.g., sensors in fields) and feeds it into a running program that generates relevant decisions or orders to control the actuators (e.g., robots or UAVs) in fields. The latter processes the data analysis tasks in a batch-wise manner. The emphasis of this type of data analysis is throughput instead of latency. A typical application scenario is to train an ML model by feeding in a large amount of data, and the training process will be running for a relatively long time. As will be discussed in Section \ref{sec:implementation}, both real-time analysis and batch processing will be run in a containerized manner, which separates the running tasks from the main logic of the system and thus provides a layer of protection of data safety.

\subsubsection{Data Storage}
\label{subsubsec: data_storage}

This layer offers a conceptual perspective of data storage within ADMA. As explained in Section \ref{subsec: concepts}, ADMA encompasses four entity modes: data, tool, model, and collection. Each entity mode in ADMA has a unique logical pathway, and cross-mode path nesting is supported. For instance, a model with the path \texttt{/user/models/model1} can include a tool with the path \texttt{/user/models/model1/tool1} nested under it. This hierarchical structuring in ADMA mirrors the organization of data in a conventional file system. Moreover, external cloud data store files can be linked to specific paths in ADMA. Additionally, ADMA maintains a clear separation between data and its associated metadata, which offers a protective layer against potential data misuse.

\subsubsection{Data Infrastructure}

This layer serves as the foundation of the entire system, functioning as the primary storage and collection point for data and including several data infrastructures, such as file systems, databases, sensors, and third-party data sources. ADMA stores user-uploaded data in a file system, where the actual data file path differs from the logical path assigned by ADMA, as detailed in Section \ref{subsubsec: data_storage}. Metadata for all data files is managed in databases, comprising a document-oriented database for storing metadata in plain text and a vector database for preserving high-dimensional vector representations of data files, as discussed in Section \ref{subsubsec: data_service}. Sensors measuring attributes such as temperature, moisture, nitrogen, and carbon dioxide can be integrated into ADMA, operating as continuous streaming data sources. Additionally, ADMA allows users to connect mainstream cloud-based data stores like Dropbox to the system, enabling authorized data folders to be mounted under specific paths.

\section{Implementation}
\label{sec:implementation}

Based on the design ideology and framework proposed in Section \ref{sec:design}, we introduce the ADMA implementation in detail.

\subsection{System Components}

\begin{figure*}[htbp]
\centerline{\includegraphics[width=0.8\linewidth]{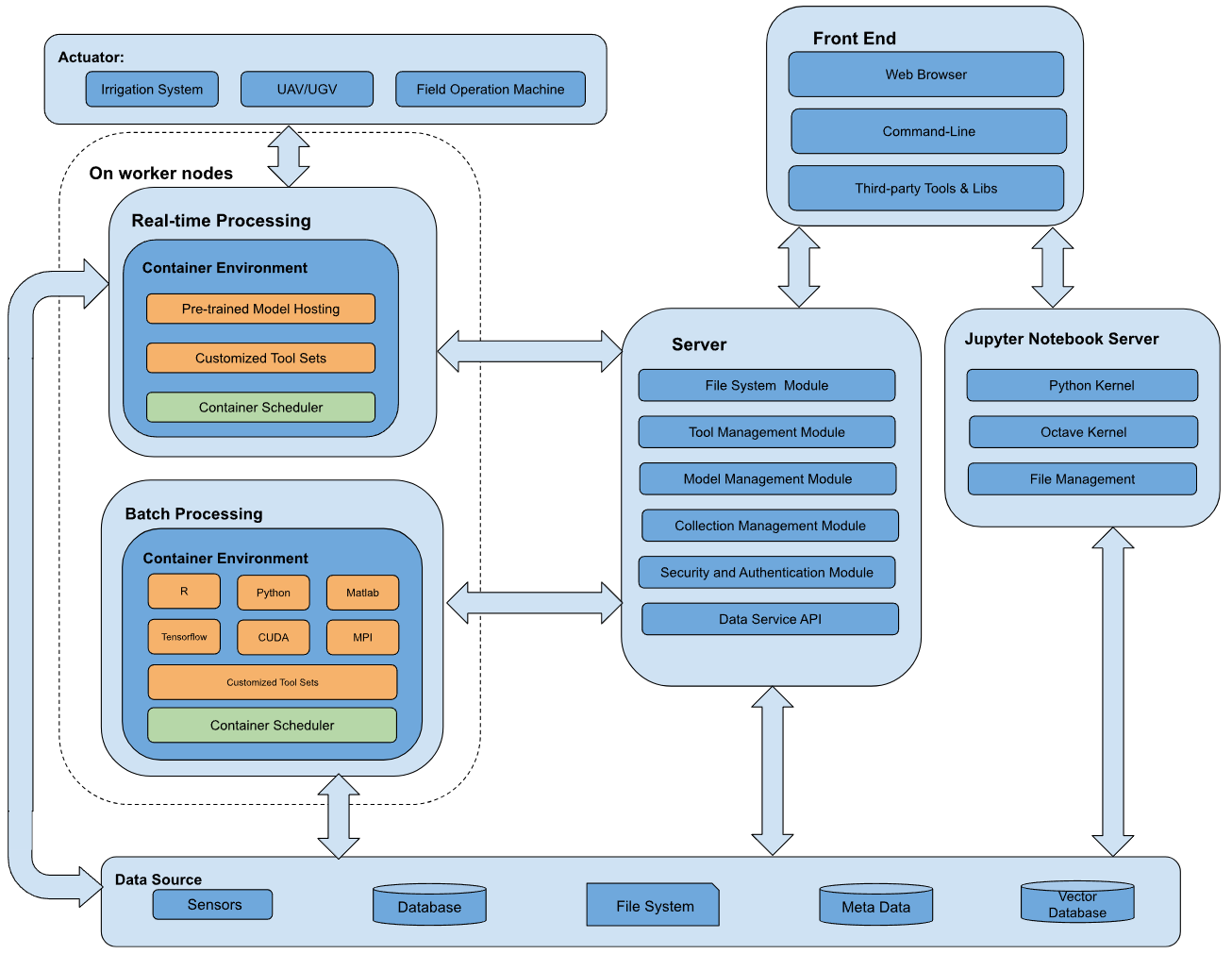}}
\caption{Main Components of ADMA}
\label{fig:components}
\end{figure*}

ADMA contains seven components: data source, batch processing, real-time processing, actuators, server, front end and JupyterHub. Figure~\ref{fig:components} illustrates the architectures of the six components.

The data source can be further categorized into 5 classes: database for storing raw data, file system for storing raw data, sensors for collecting real-time data ,database for storing metadata and vector database for storing high dimensional vector representations of data.

The batch processing component is the engine of the whole system. The containerized environment can host various types of containerized tools such as R, Python, Matlab, Tensorflow, Cuda, MPI, and many more customized tools for the specific disciplines. There is a  container scheduler such as Swarm or Kubernetes to manage the running of containers. Various user-defined data processing and model training can be conducted in the container environment. The output of the batch processing will be stored back to the data source for further retrieval and processing.

The real-time processing component connects to the sensors and uses pipes to stream the real-time data for further processing. Similar to batch processing, a containerized environment is also supported, where hosted pre-trained models and customized tools can process data and make decisions in a real-time or near real-time fashion.  The real-time analytical results can be fed to the data processing module and data service API inside the server component. The decisions and commands can be output to the actuator to execute. This component is a necessary complement to batch processing when real-time analysis and control are required, and relatively low accuracy is tolerated.

The actuator is included in the system to make it more like a fully functioning control system. Connected to the real-time processing component, the actuator executes the commands/decisions output by the hosted models or user-coded tools. The types of actuators vary among different disciplines. For biological system engineering, to name a few, there are actuators such as irrigation systems, robotics, and field operation machines. 

\begin{figure}[t]
    \centering
    \begin{subfigure}{0.4\textwidth}
        \centering
        \includegraphics[width=\linewidth]{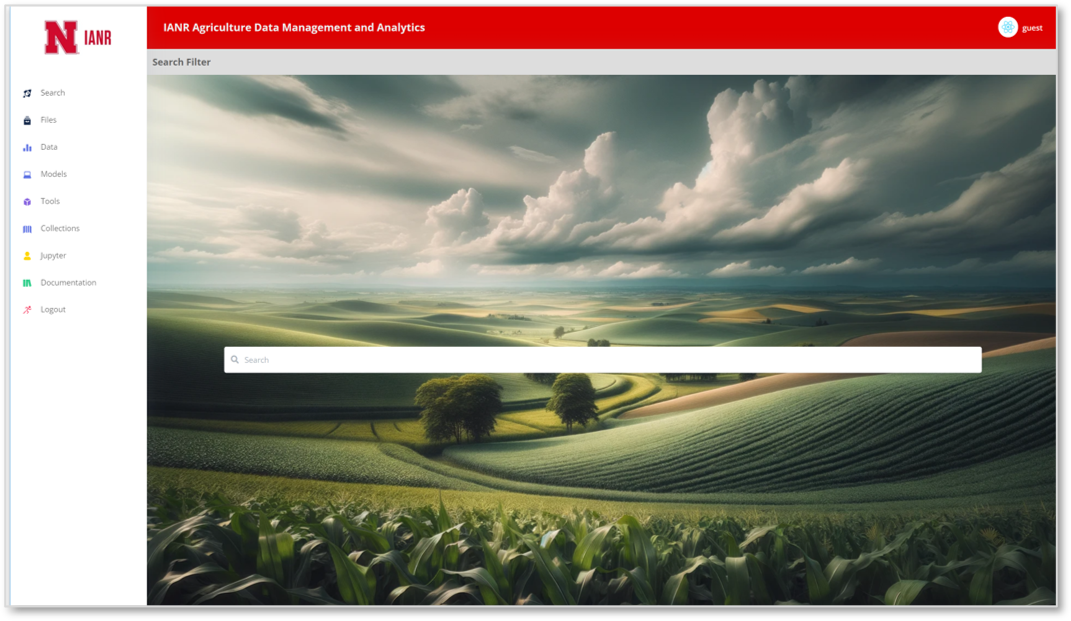}
        \caption{Search Box}
        \label{fig:data_search_1}
    \end{subfigure}%
    \hfill
    \begin{subfigure}{0.4\textwidth}
        \centering
        \includegraphics[width=\linewidth]{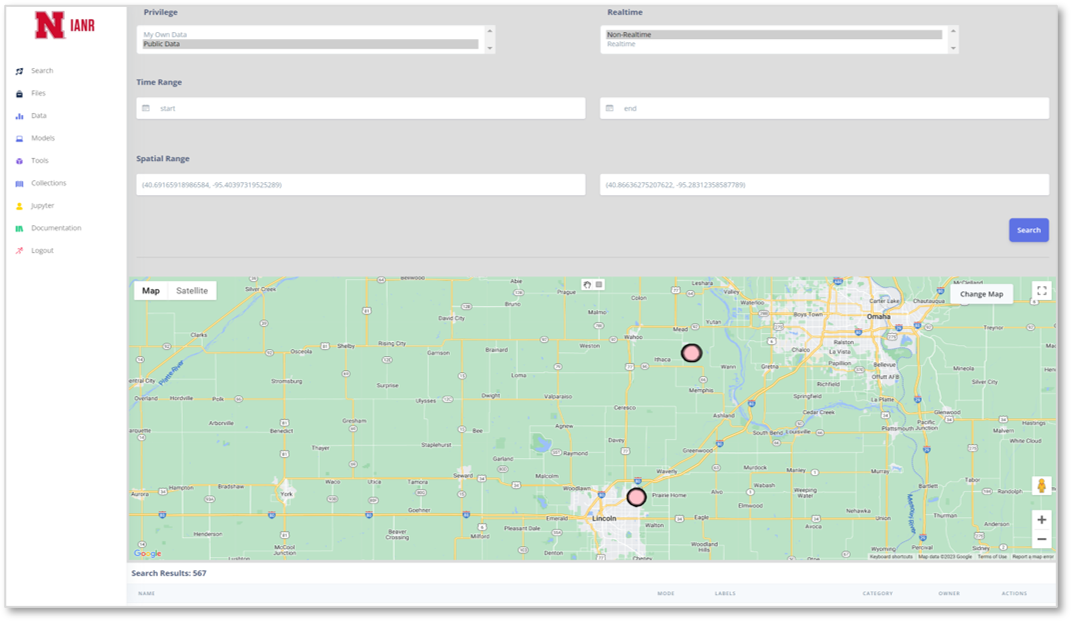}
        \caption{Map Visualization}
        \label{fig:data_search_2}
    \end{subfigure}%
    \hfill
    \begin{subfigure}{0.4\textwidth}
        \centering
        \includegraphics[width=\linewidth]{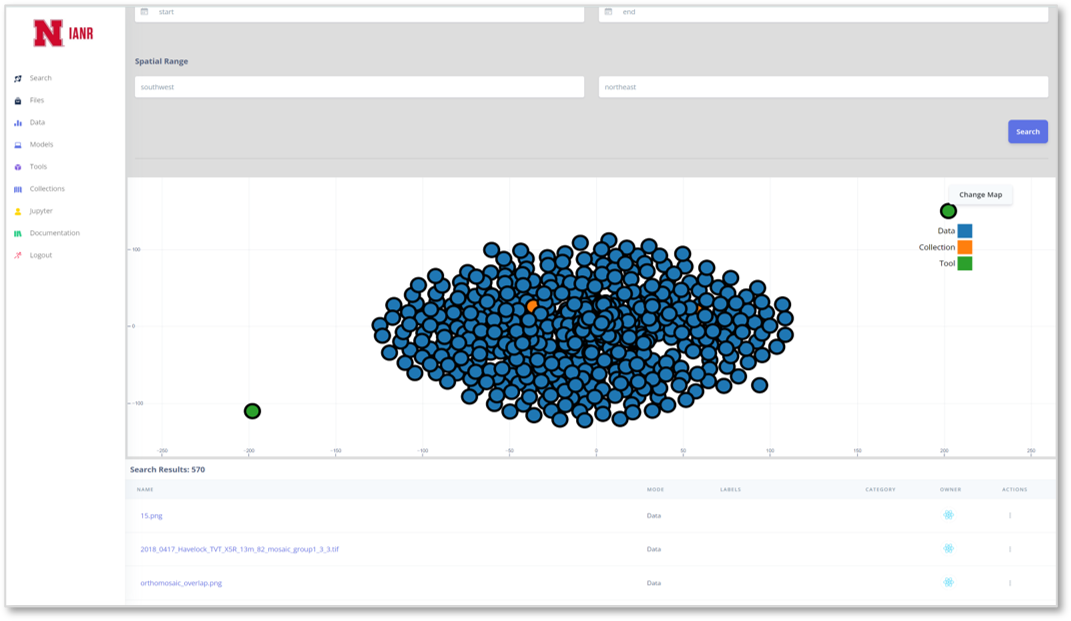}
        \caption{Embedding Visualization}
        \label{fig:data_search_3}
    \end{subfigure}%
    \hfill
    \begin{subfigure}{0.4\textwidth}
        \centering
        \includegraphics[width=\linewidth]{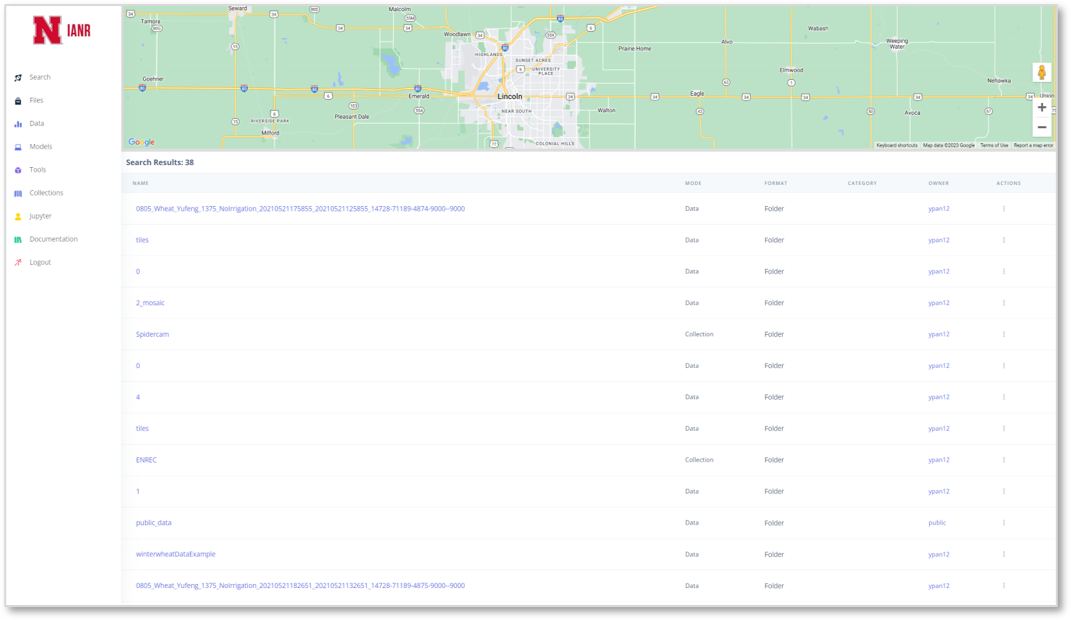}
        \caption{Search Results List}
        \label{fig:data_search_4}
    \end{subfigure}%
    \hfill   

    \caption{Data Search}
    \label{fig:data_search}
\end{figure}

The server is the operational handler of the whole system. The data management module is responsible for ordinary operations such as file search, retrieve, upload, delete, update, and processing. The tool management module keeps the information of the tool sets used in the system. The model management module is responsible for maintaining all the pre-trained models. The data processing module talks to and manages the batch processing and real-time processing components. This module is an umbrella module for all the data processing jobs, such as data pre-processing, data co-locating, batch processing, model training,  model hosting, real-time processing, etc. The security and authentication module guarantees the data storage and transmission is secure, and each user has separate access permission and view of the data. Data service API acts as the portal for the front end to manipulate the data in our system.

The front end runs on the user-side machine and can be a web browser, command-line, or third-party software. The front end communicates with the server through the data service API. Users of our system can conduct ordinary file management operations and initiate data processing requests to the system. The analytical results can be displayed in the front end for the user to view.

To provide the users with more flexibility, ADMA connects with JupyterHub. Each user can create their own JupyterLab, which is the latest web-based interactive development environment for notebooks, code, and data in the field of data science, scientific computing, computational journalism, and machine learning. User can utilize the same workspace to put their data and tools in both ADMA and JupyterLab, in this way the user can shift between two systems seamlessly.


\section{Evaluation}
\label{sec:evaluation}

\subsection{System Demo}
\subsubsection{Search}

ADMA supports semantic search by natural language (Figure \ref{fig:data_search_1}). The embedding of query sentences will be compared with the embeddings of each file's metadata, and the most similar results will be retrieved and displayed. There is also a filter supporting search by various criteria such as category, mode, format, label, privilege, real-time/non-real-time, time range, and spatial range. Search results will be displayed on a map based on their locations (Figure \ref{fig:data_search_2}) or visualized on a 2D space based on the embeddings of their metadata (Figure \ref{fig:data_search_3}), where each dot represents a file or folder and each color denotes a mode. Search results will also be displayed in a list, a conventional way used by a typical query system (Figure \ref{fig:data_search_4}).

\subsubsection{File Management}

\begin{figure}[!t]
    \centering
    \begin{subfigure}{0.4\textwidth}
        \centering
        \includegraphics[width=\linewidth]{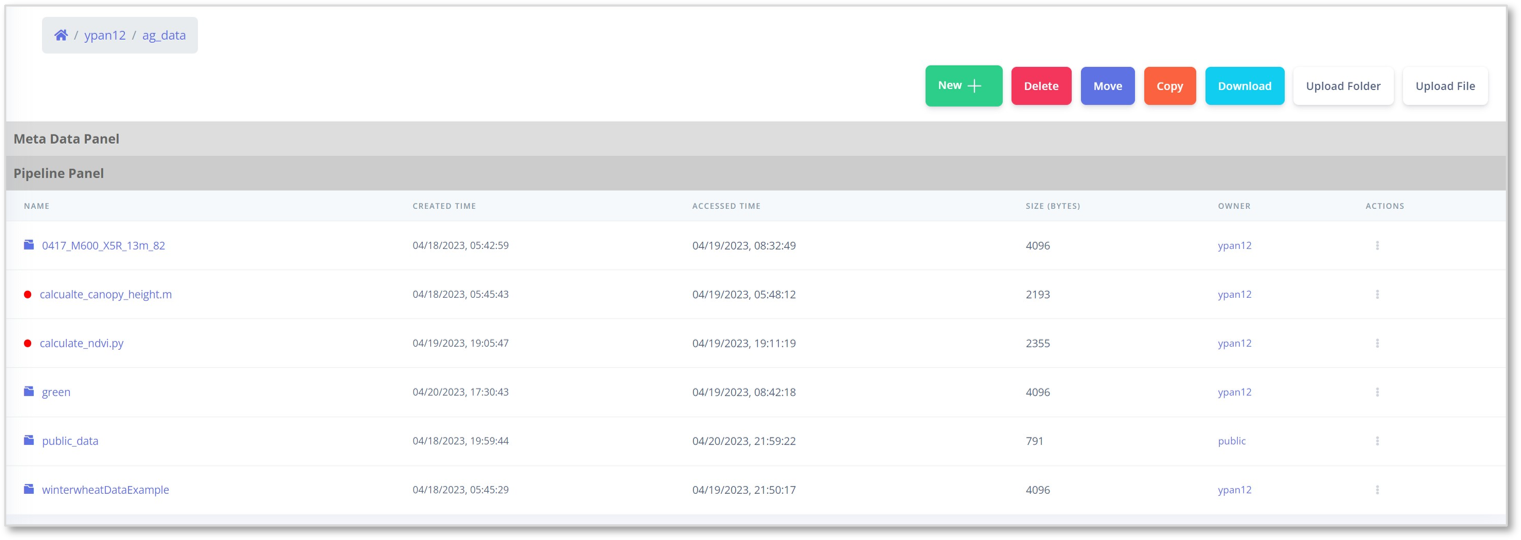}
        \caption{Basic File Management Operations}
        \label{fig:file_management_1}
    \end{subfigure}%
    \hfill
    \begin{subfigure}{0.4\textwidth}
        \centering
        \includegraphics[width=\linewidth]{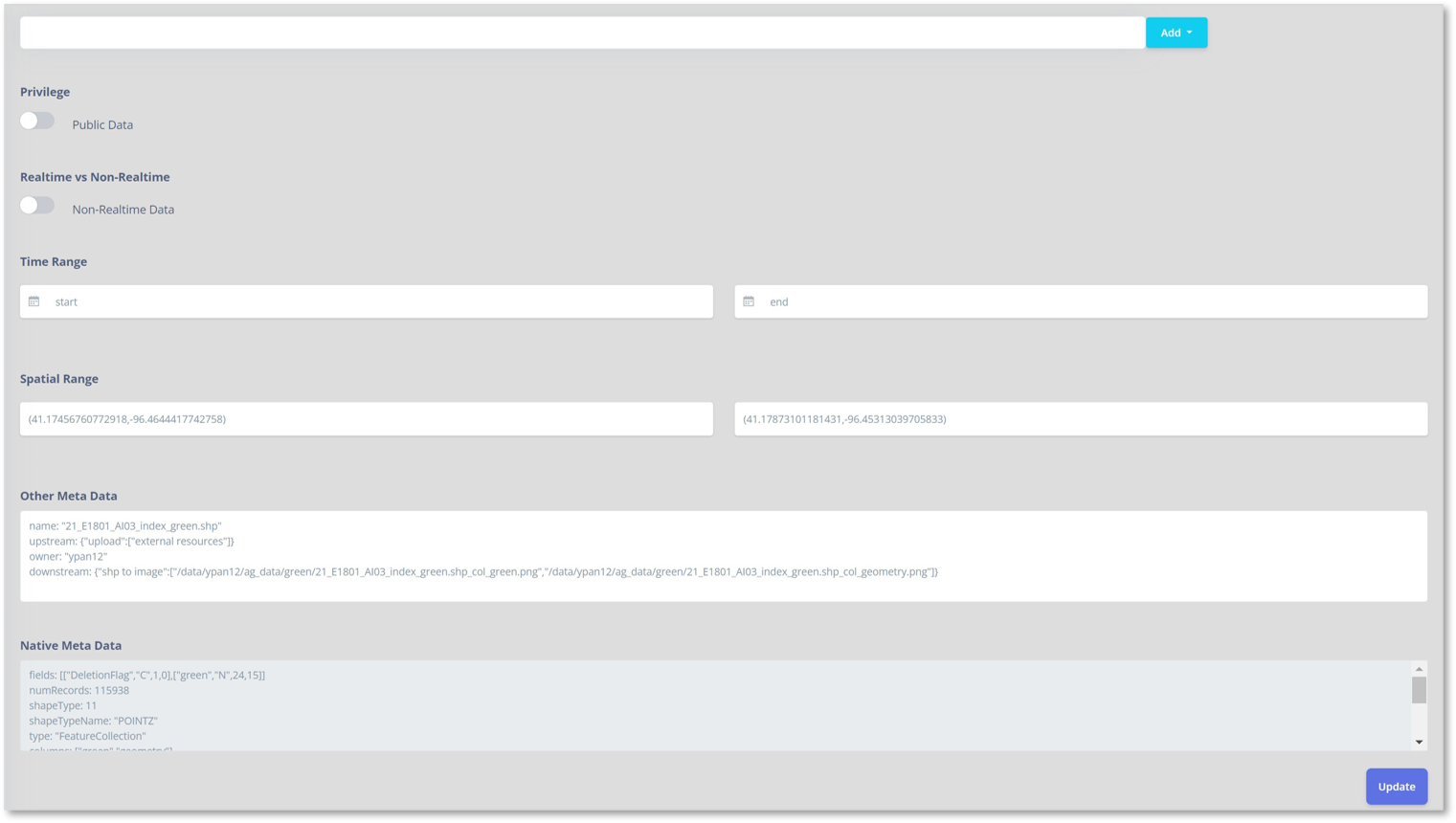}
        \caption{Meta Data Panel}
        \label{fig:file_management_2}
    \end{subfigure}%
    \hfill
    \caption{File Management on ADMA}
    \label{fig:file_management}
\end{figure}

The file management module in ADMA supports basic operations on files, such as creating, uploading, deleting, moving, copying, downloading, and editing (Figure \ref{fig:file_management_1}). Each user will be assigned a root dir named by their username \texttt{/username}, and all the data will be put into the directory \texttt{/username/ag\_data}. Each user will also see a folder named \texttt{public\_data} under \texttt{/username/ag\_data}, which contains all the public data from other users. ADMA keeps a metadata panel for each folder or file, which can be used to update various metadata items (Figure \ref{fig:file_management_2}). For user-uploaded programs or scripts, the system presents a tool panel for executing tools, and we will introduce it in Secion \ref{subsubsec:running_tools}. Each folder or file also has a pipeline panel, which will be introduced in Section \ref{subsubsec:pipeline}.

\subsubsection{Data Rendering and Online Editing}

\begin{figure}[!t]
    \centering
    \begin{subfigure}{0.4\textwidth}
        \centering
        \includegraphics[width=\linewidth]{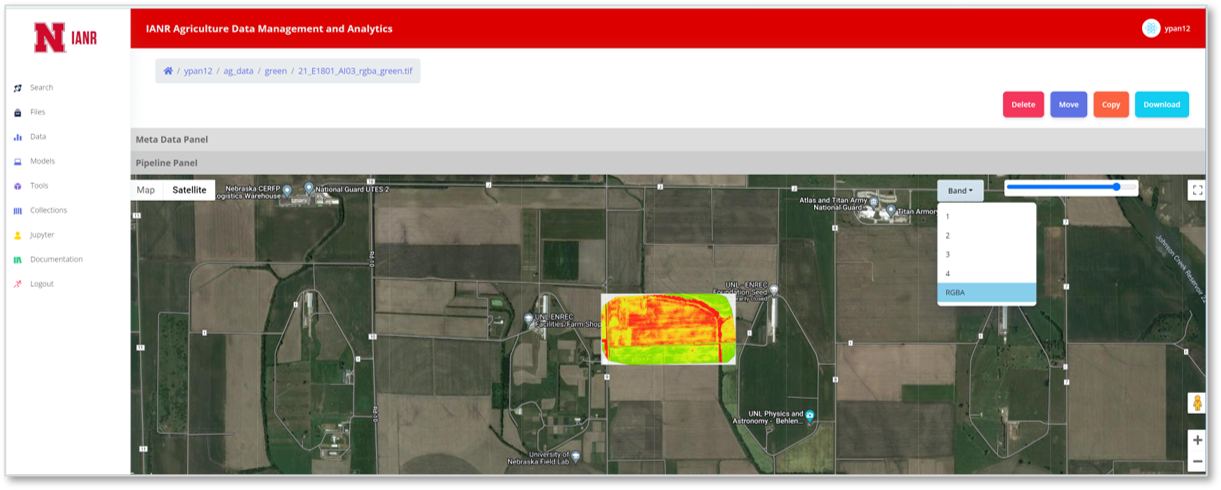}
        \caption{Rendering .tif File on Map}
        \label{fig:data_rendering_1}
    \end{subfigure}%
    \hfill
    \begin{subfigure}{0.4\textwidth}
        \centering
        \includegraphics[width=\linewidth]{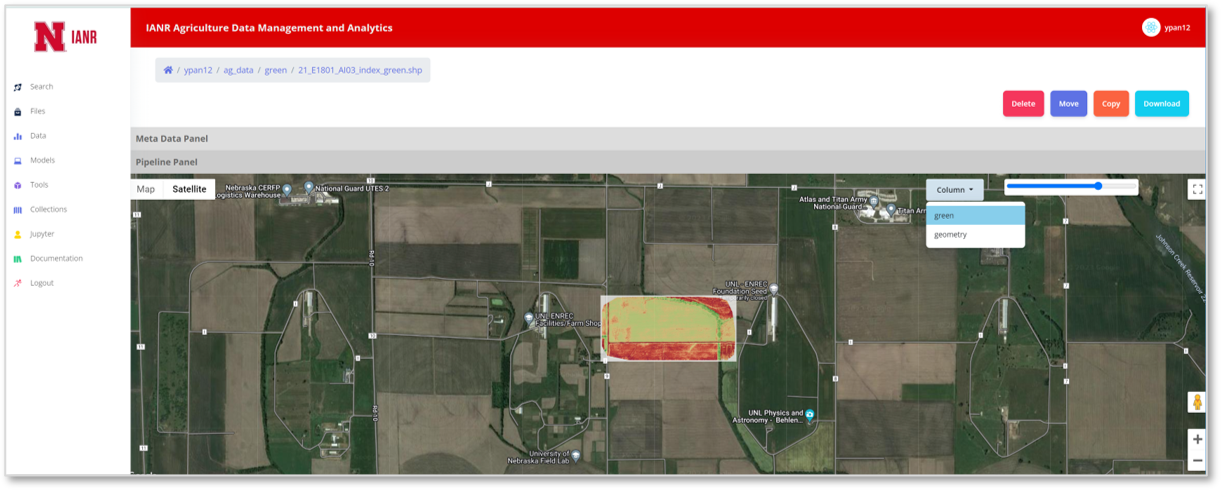}
        \caption{Meta Data Panel}
        \label{fig:data_rendering_2}
    \end{subfigure}%
    \hfill
    \begin{subfigure}{0.4\textwidth}
        \centering
        \includegraphics[width=\linewidth]{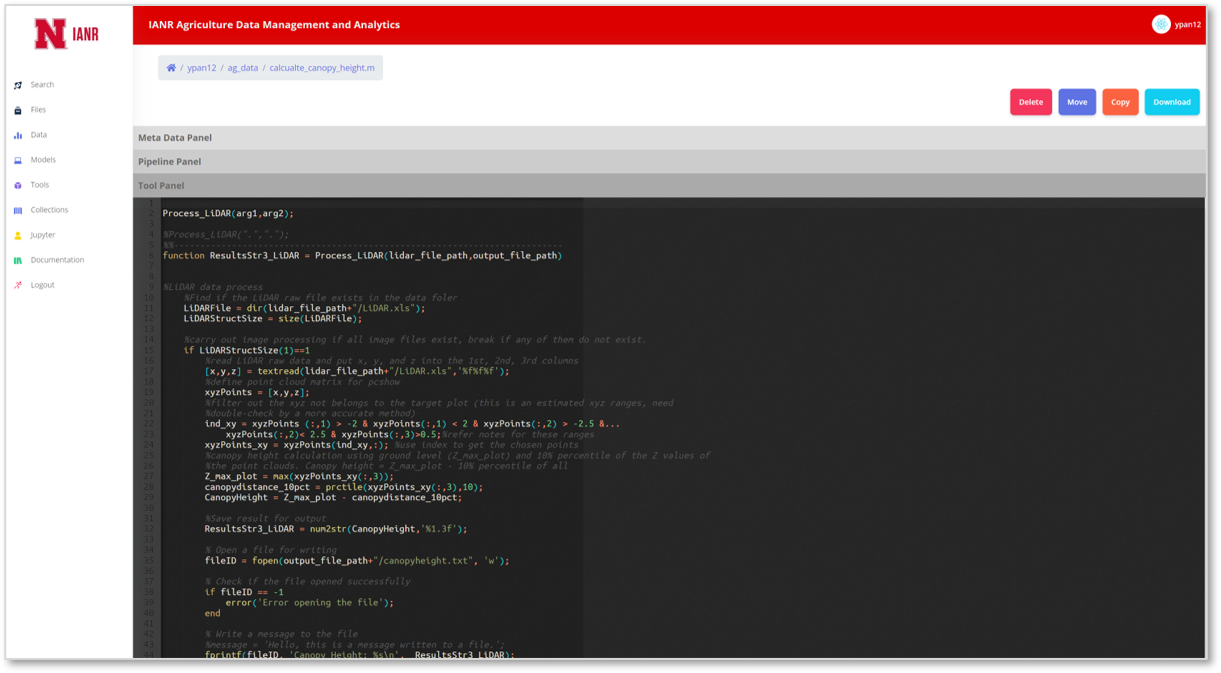}
        \caption{Online Matlab Code Editing}
        \label{fig:data_rendering_3}
    \end{subfigure}%
    \hfill
    \begin{subfigure}{0.4\textwidth}
        \centering
        \includegraphics[width=\linewidth]{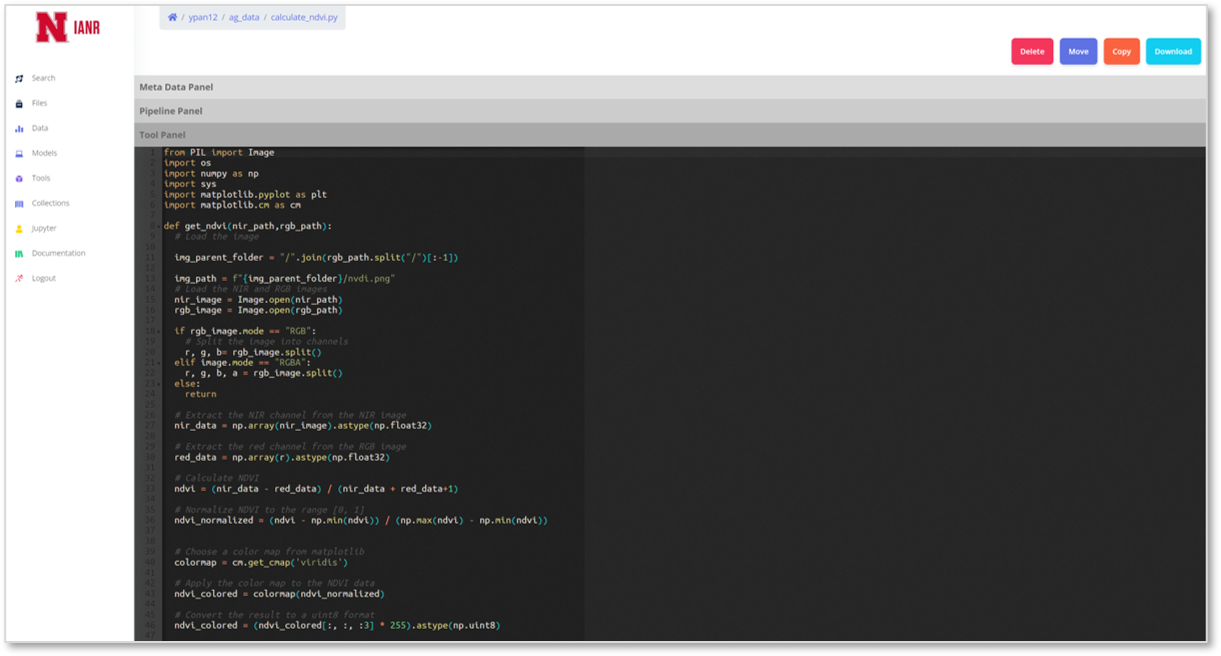}
        \caption{Online Python Code Editing}
        \label{fig:data_rendering_4}
    \end{subfigure}%
    \hfill

    \caption{Data Rendering and Online Editing on ADMA}
    \label{fig:data_rendering}
\end{figure}

ADMA provides native supports for rendering or editing some frequently used data formats, such as .tif (Figure \ref{fig:data_rendering_1}), .shp files (Figure \ref{fig:data_rendering_2}), scripting languages such as python (Figure \ref{fig:data_rendering_4}), and R and Matlab (Figure \ref{fig:data_rendering_3}). The geoinformation required to render .tif and .shp files are extracted from the native metadata of the file themselves.

\subsubsection{Running Tools}
\label{subsubsec:running_tools}

\begin{figure}[!t]
    \centering
    \begin{subfigure}{0.4\textwidth}
        \centering
        \includegraphics[width=\linewidth]{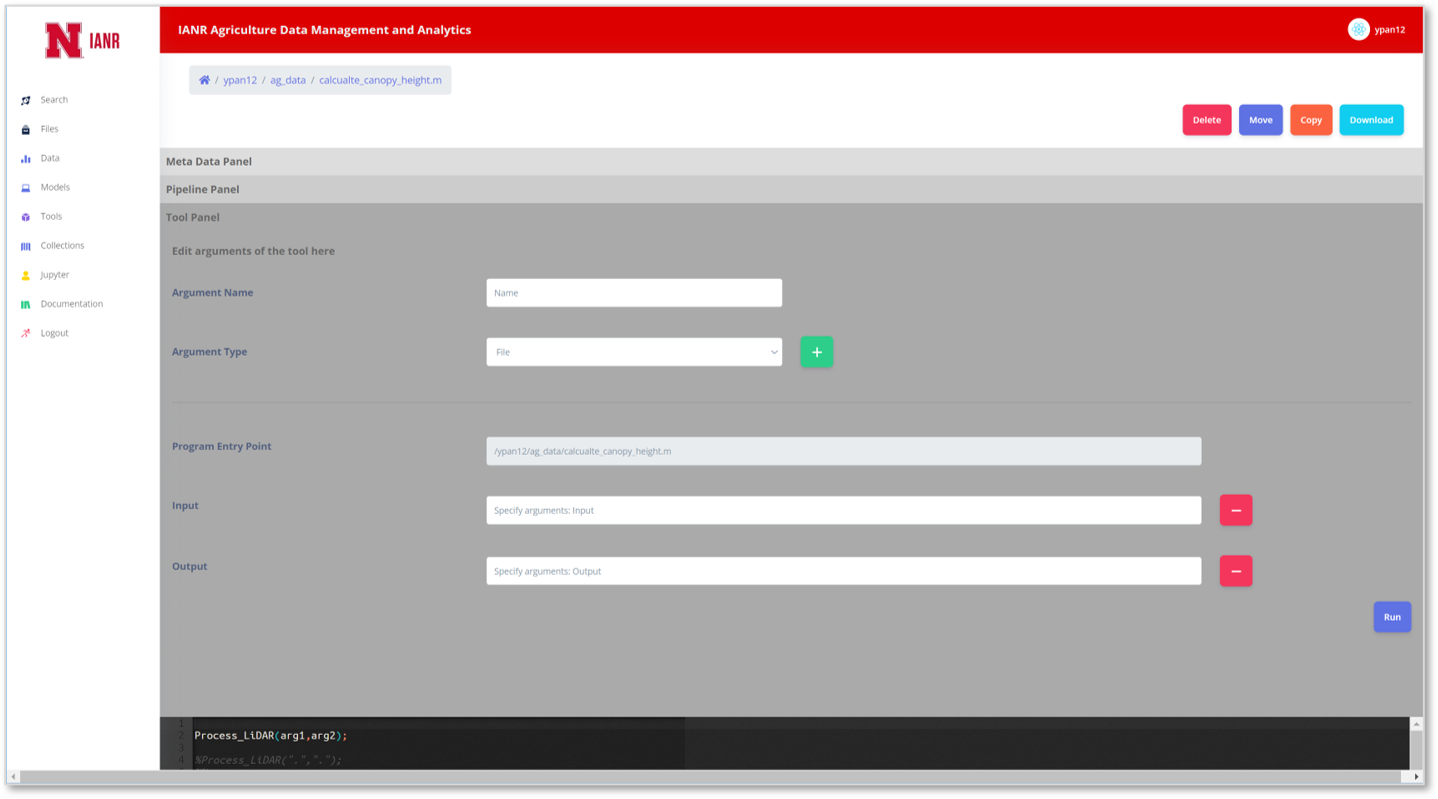}
        \caption{Tool Panel}
        \label{fig:running_tools_1}
    \end{subfigure}%
    \hfill
    \begin{subfigure}{0.4\textwidth}
        \centering
        \includegraphics[width=\linewidth]{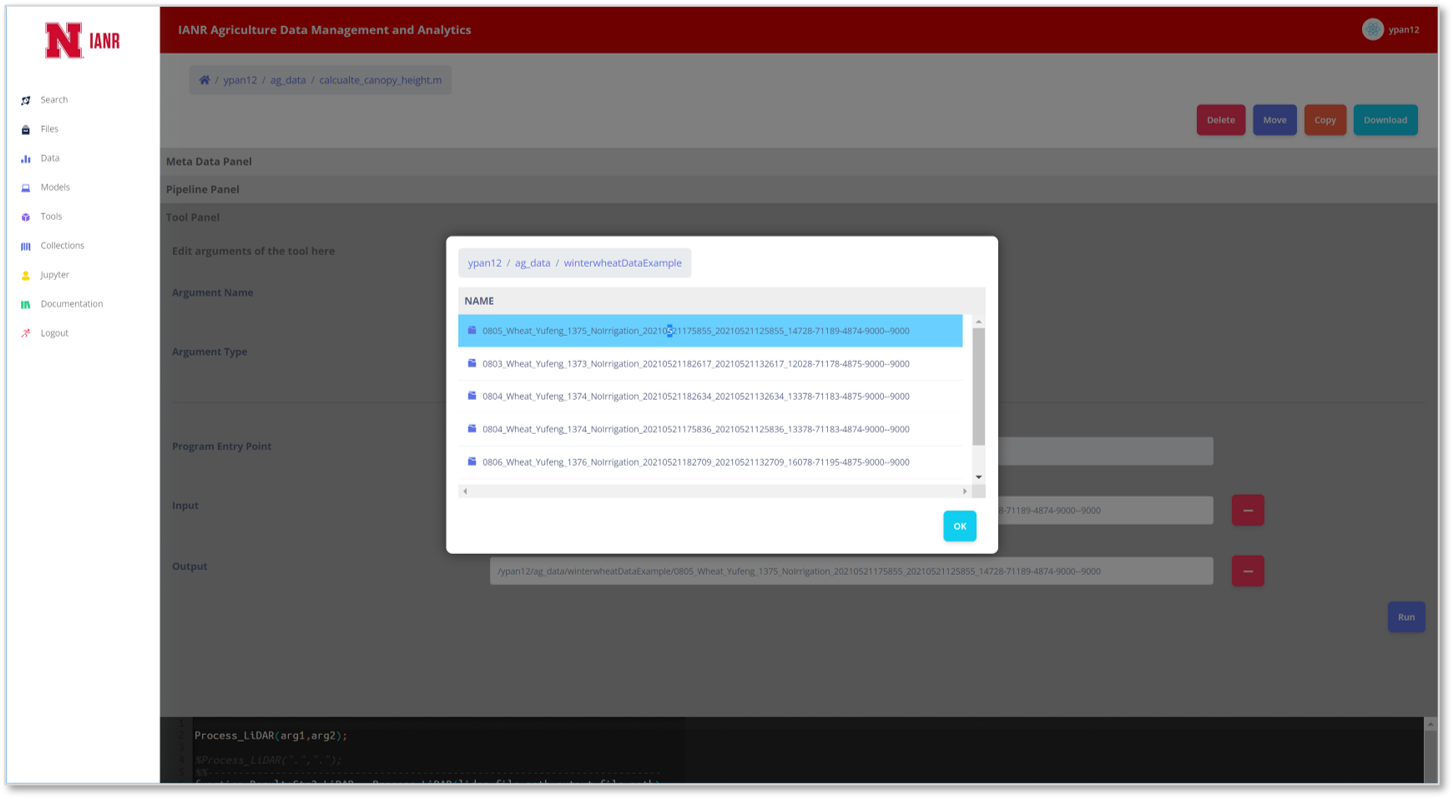}
        \caption{File Selection}
        \label{fig:running_tools_2}
    \end{subfigure}%
    \hfill
    \begin{subfigure}{0.4\textwidth}
        \centering
        \includegraphics[width=\linewidth]{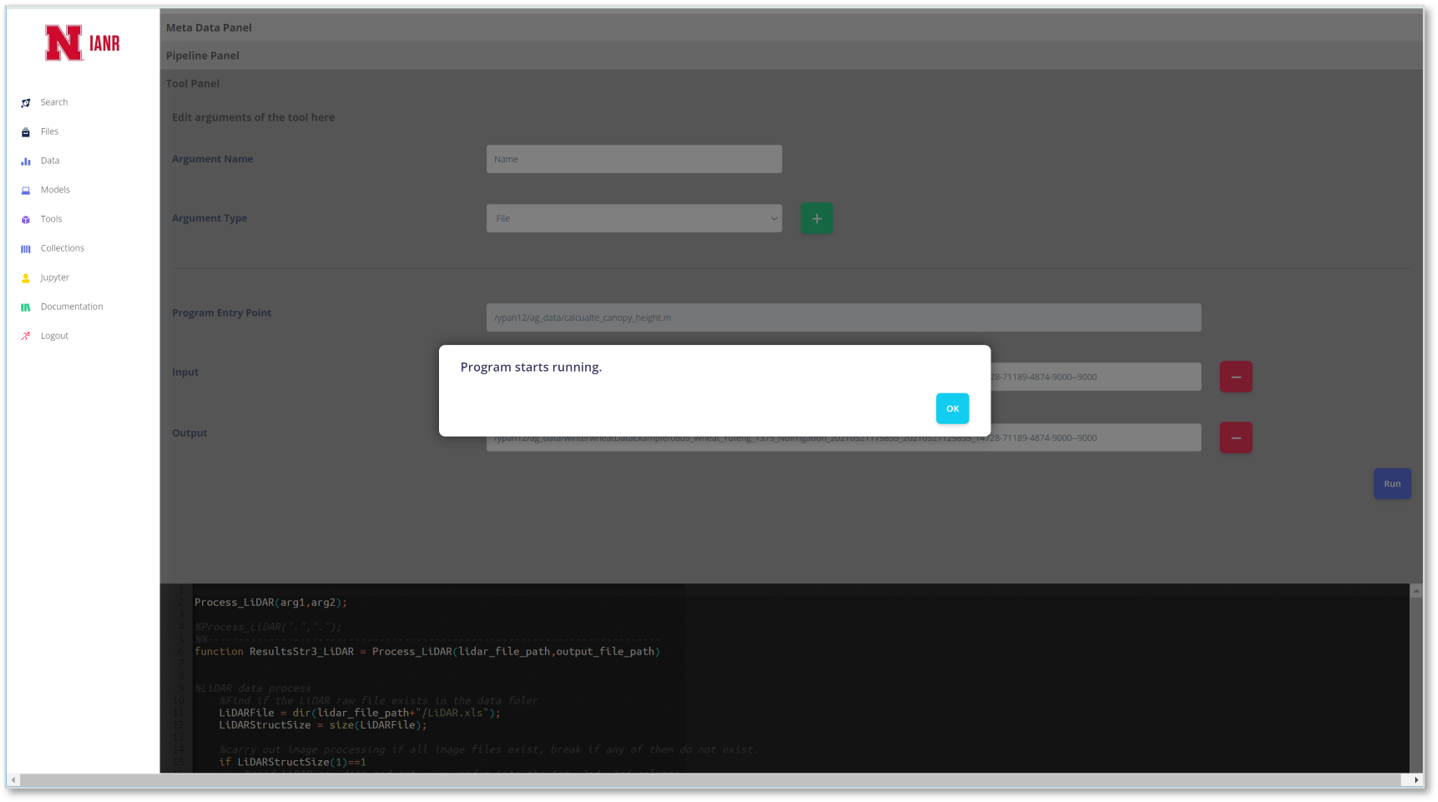}
        \caption{Execution Starting}
        \label{fig:running_tools_3}
    \end{subfigure}%
    \hfill
    \begin{subfigure}{0.4\textwidth}
        \centering
        \includegraphics[width=\linewidth]{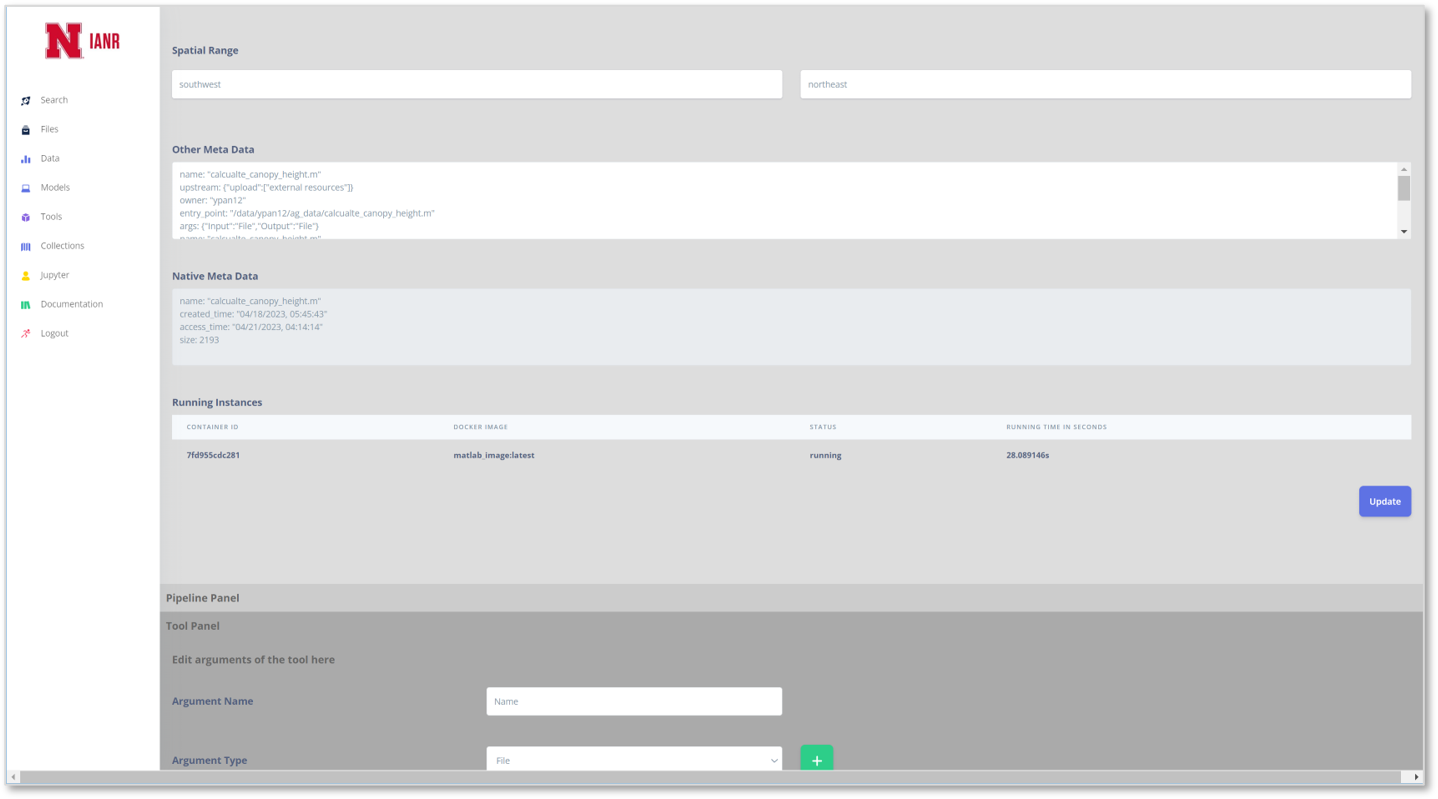}
        \caption{Running Program Instances}
        \label{fig:running_tools_4}
    \end{subfigure}%
    \hfill

    \caption{Running Tools on ADMA}
    \label{fig:data_rendering}
\end{figure}

For each user-uploaded tool, ADMA displays a tool panel in which the user can edit, add, or remove command-line arguments of the tool (Figure \ref{fig:running_tools_1}). If the argument is a path, ADMA allows users to select files in their \texttt{ag\_data} folder (Figure \ref{fig:running_tools_2}). Once the Run button is clicked, a message will be prompted indicating the program starts running in a containerized environment (Figure \ref{fig:running_tools_3}). Then, a user can check the execution status of the program by scrolling to the bottom of the metadata panel, where the container ID, docker image, status, and running time of the running instances of the tool will be displayed (Figure \ref{fig:running_tools_4}). While the tool is running, there will also be a green dot before the name of the tool in the file management system. After the program stops, users can check the output in their specified location.

\subsubsection{Pipeline Management}
\label{subsubsec:pipeline}

\begin{figure}[!t]
    \centering
    \begin{subfigure}{0.4\textwidth}
        \centering
        \includegraphics[width=\linewidth]{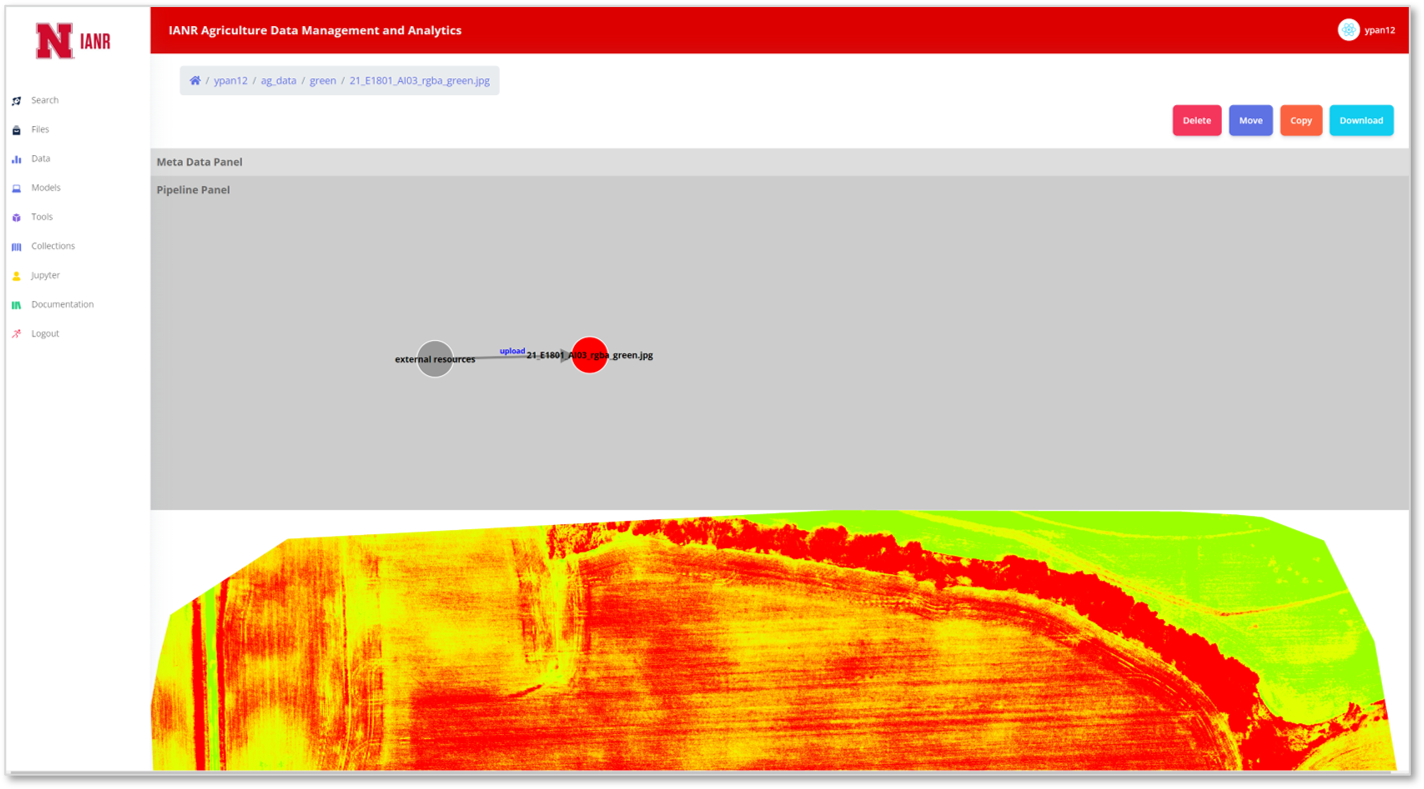}
        \caption{Pipeline of Uploading a File}
        \label{fig:pipeline_1}
    \end{subfigure}%
    \hfill
    \begin{subfigure}{0.4\textwidth}
        \centering
        \includegraphics[width=\linewidth]{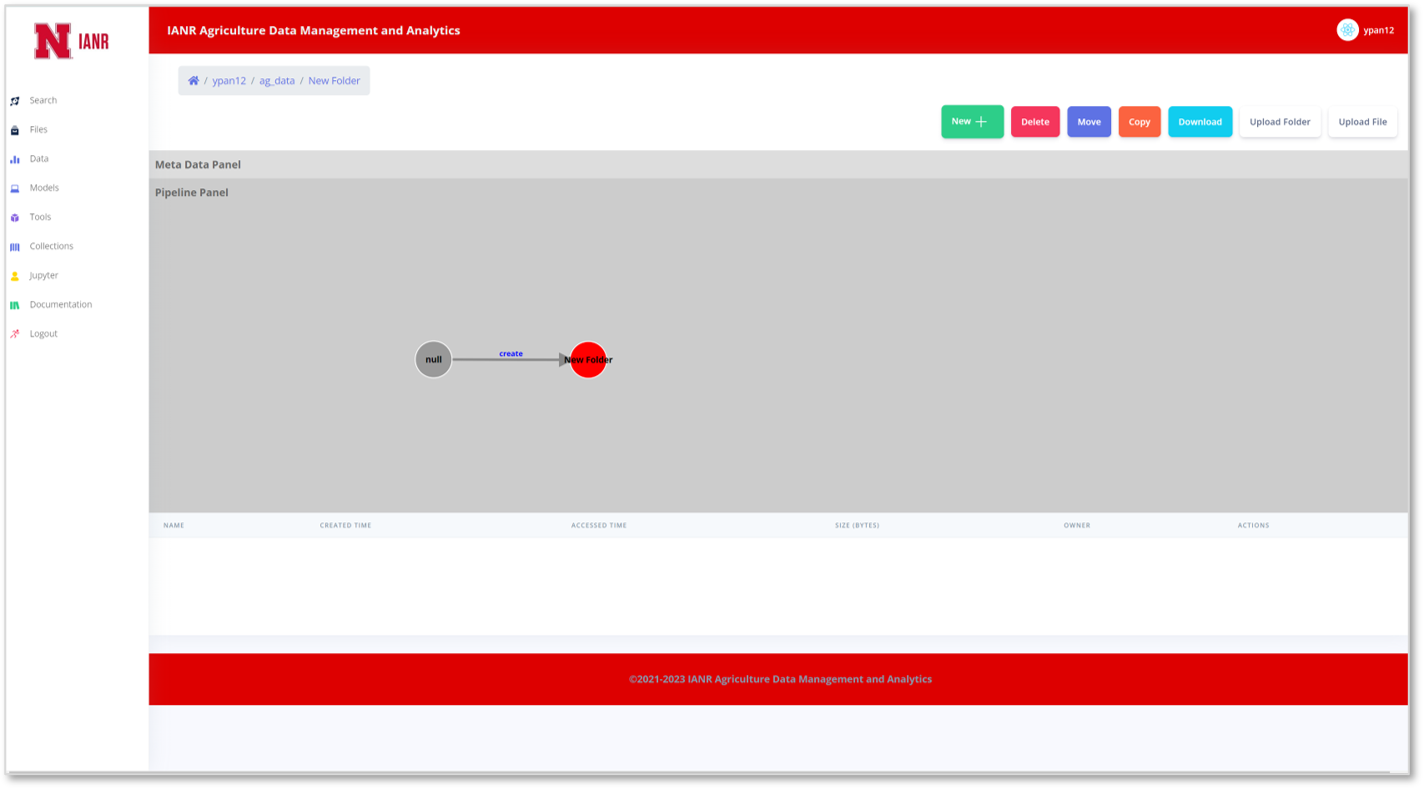}
        \caption{Pipeline of Creating a Folder}
        \label{fig:pipeline_2}
    \end{subfigure}%
    \hfill
    \begin{subfigure}{0.4\textwidth}
        \centering
        \includegraphics[width=\linewidth]{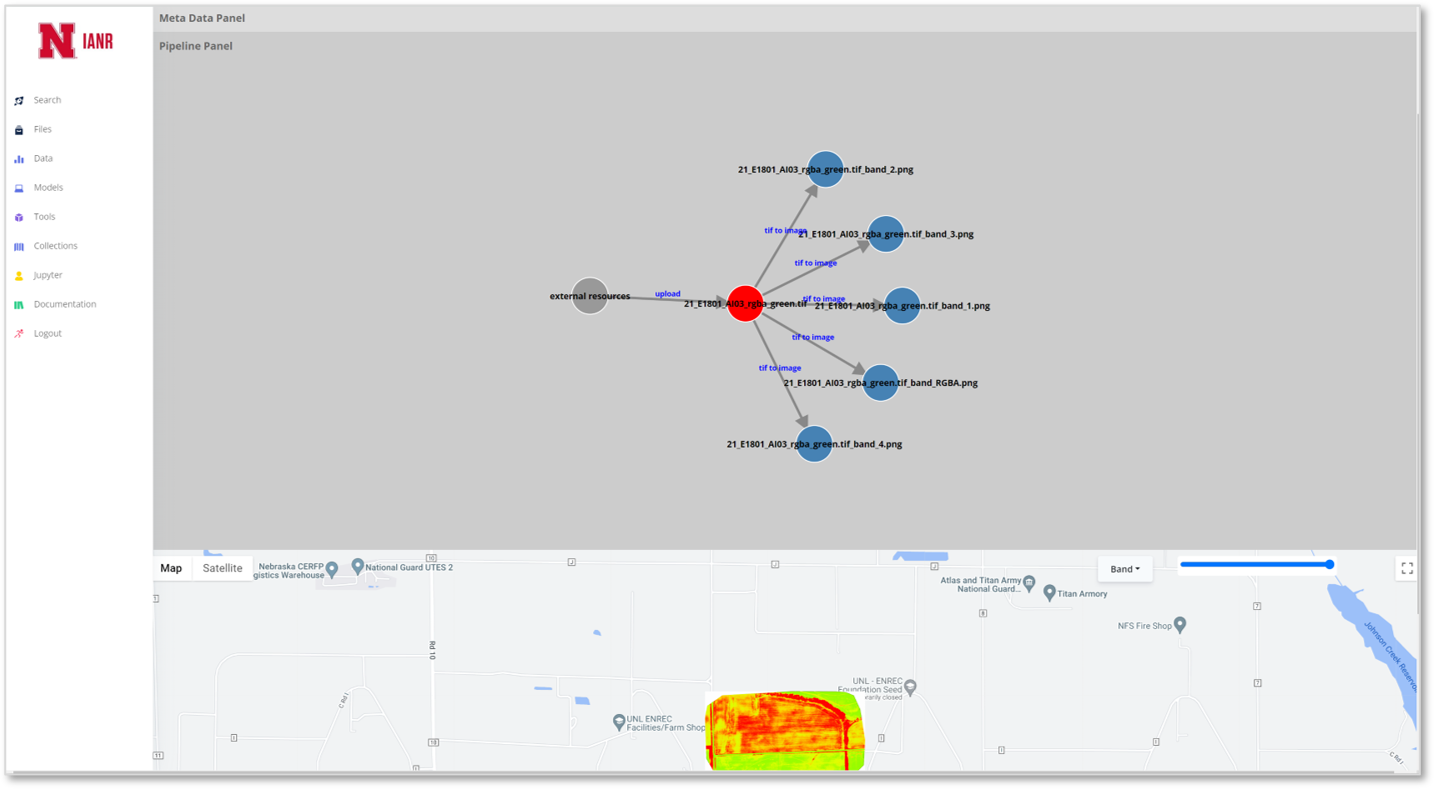}
        \caption{Pipeline of Converting TIF to PNG}
        \label{fig:pipeline_3}
    \end{subfigure}%
    \hfill
    \begin{subfigure}{0.4\textwidth}
        \centering
        \includegraphics[width=\linewidth]{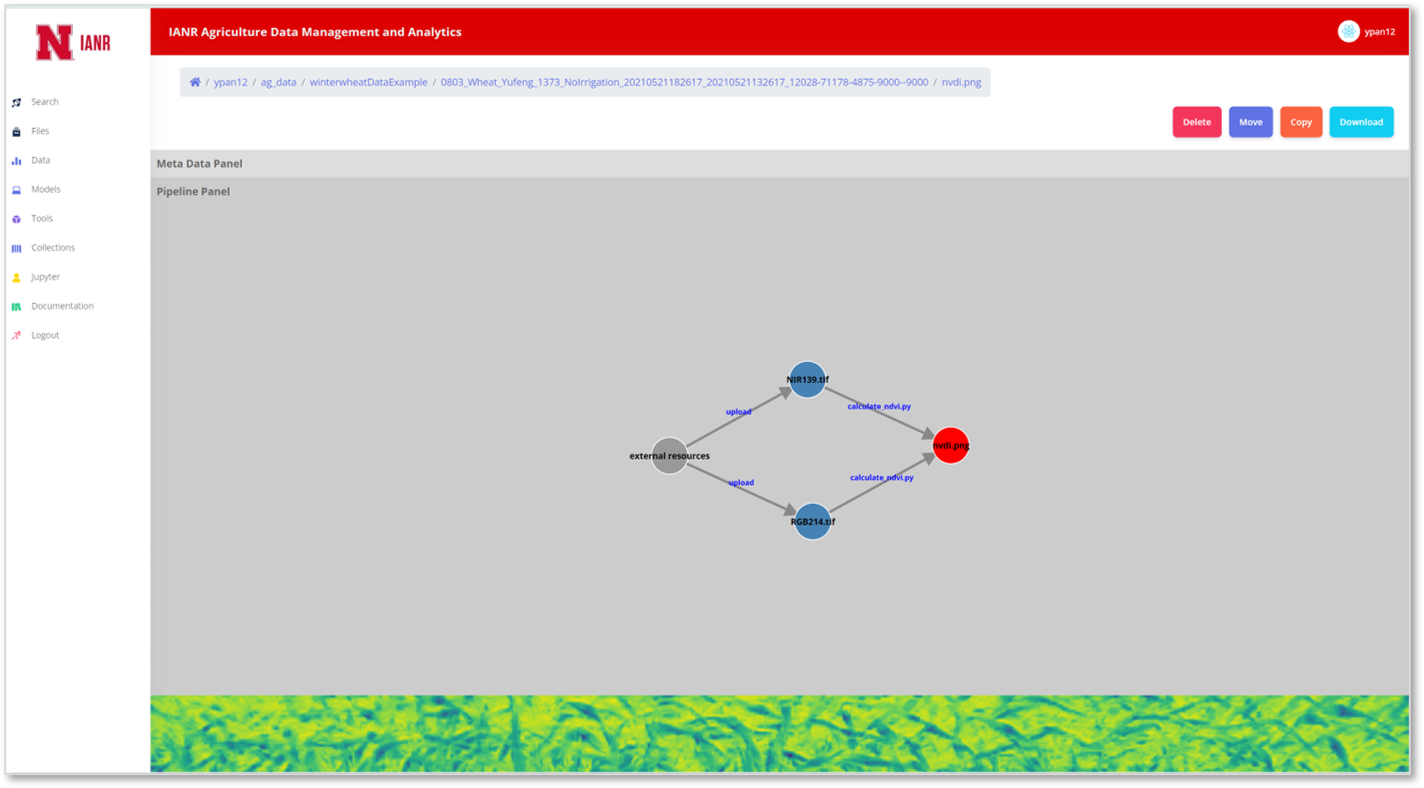}
        \caption{Pipeline of User Uploaded Tool}
        \label{fig:pipeline_4}
    \end{subfigure}%
    \hfill
    \caption{Pipeline Management on ADMA}
    \label{fig:data_rendering}
\end{figure}

For each file, ADMA automatically keeps track of its history and keeps a pipeline panel illustrating of the file together with the upstream and downstream files and corresponding operations of each step. ADMA keeps track of all native operations and user-uploaded tools. For instance, in Figure \ref{fig:pipeline_1}, when a user uploads a file named \texttt{21\_E1801\_A103\_rgba\_green.png}, the pipeline will show it is uploaded from an external resource. In Figure \ref{fig:pipeline_2}, when the user creates a new folder, the pipeline will indicate it is created from \texttt{null}. Figure \ref{fig:pipeline_3} shows the pipeline of a user-uploaded TIF file being converted to a PNG file for each of its bands. Figure \ref{fig:pipeline_4} illustrates how a user-uploaded tool: \texttt{calculate\_ndvi.py} converts two user-uploaded TIF files and generates one image file \texttt{ndvi.png}.

\subsubsection{Collections}

\begin{figure}[!t]
    \centering
    \begin{subfigure}{0.4\textwidth}
        \centering
        \includegraphics[width=\linewidth]{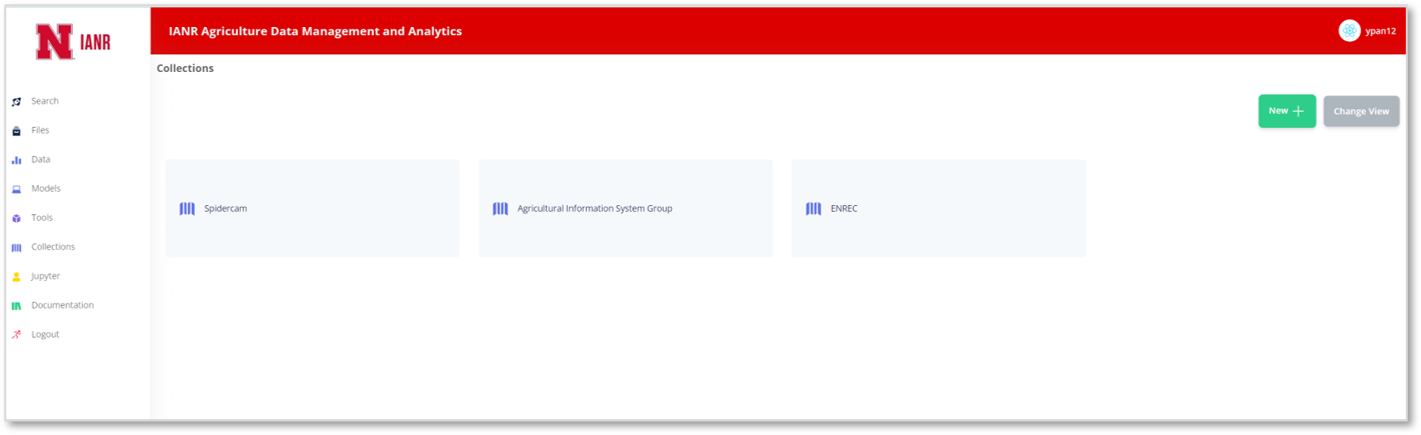}
        \caption{Page of Collections}
        \label{fig:collections_1}
    \end{subfigure}%
    \hfill
    \begin{subfigure}{0.4\textwidth}
        \centering
        \includegraphics[width=\linewidth]{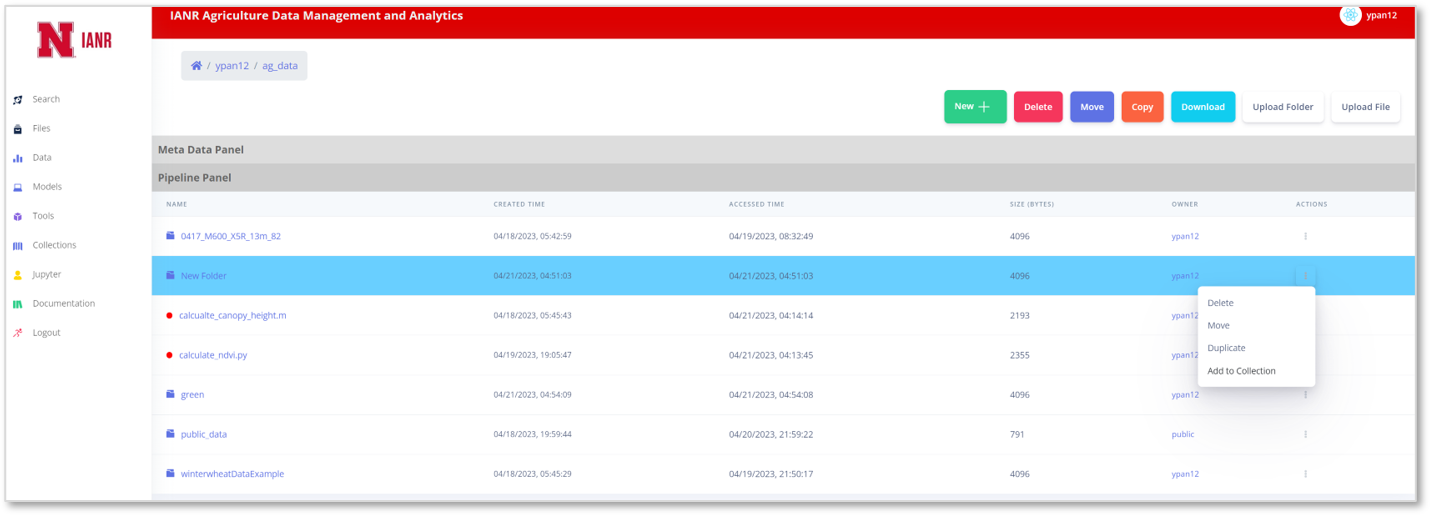}
        \caption{Add Data to Collection}
        \label{fig:collections_2}
    \end{subfigure}%

    \caption{Collection Management on ADMA}
    \label{fig:data_rendering}
\end{figure}

Besides original file hierarchies, collections provide another way to organize data. ADMA allows a user to create a new collection (Figure \ref{fig:collections_1}) and add their data to the collection (Figure \ref{fig:collections_2}). Collection is also one of the four modes in ADMA, besides data, tool, and model, which means a collection is searchable, and each collection has a metadata page.

\subsubsection{Model Training and Hosting}

ADMA supports the training and hosting of machine learning models. The process of model training is no different from the one described in running tools in Section \ref{subsubsec:running_tools}, except that here, we invoke a model training program and specify the required dataset using the tool panel. The training process will also be running in a separate container. After the model is trained, the files containing the parameters of the model will be stored in a user-specified location. Then, during inference, the user can invoke the model inferring program by using the tool panel of the folder of the trained model.

\subsubsection{Jupyter}

To provide users with more flexibility, ADMA connects with JupyterHub. Each user can create their own JupyterLab, which is the latest web-based interactive development environment for notebooks, code, and data in the fields of data science, scientific computing, computational journalism, and machine learning. Users can utilize the same workspace to put their data and tools in both ADMA and JupyterLab, and in this way, users can shift between two systems seamlessly. Data privacy and safety are protected by running JupyterLab for each user in a containerized environment. Note that ADMA can not keep track of the operations conducted through JupyterLab, which means the pipeline for any operations in JupyterLab will not be automatically generated. 

\subsubsection{Data API}

\begin{figure}[!t]
    \centering
    \begin{subfigure}{0.4\textwidth}
        \centering
        \includegraphics[width=\linewidth]{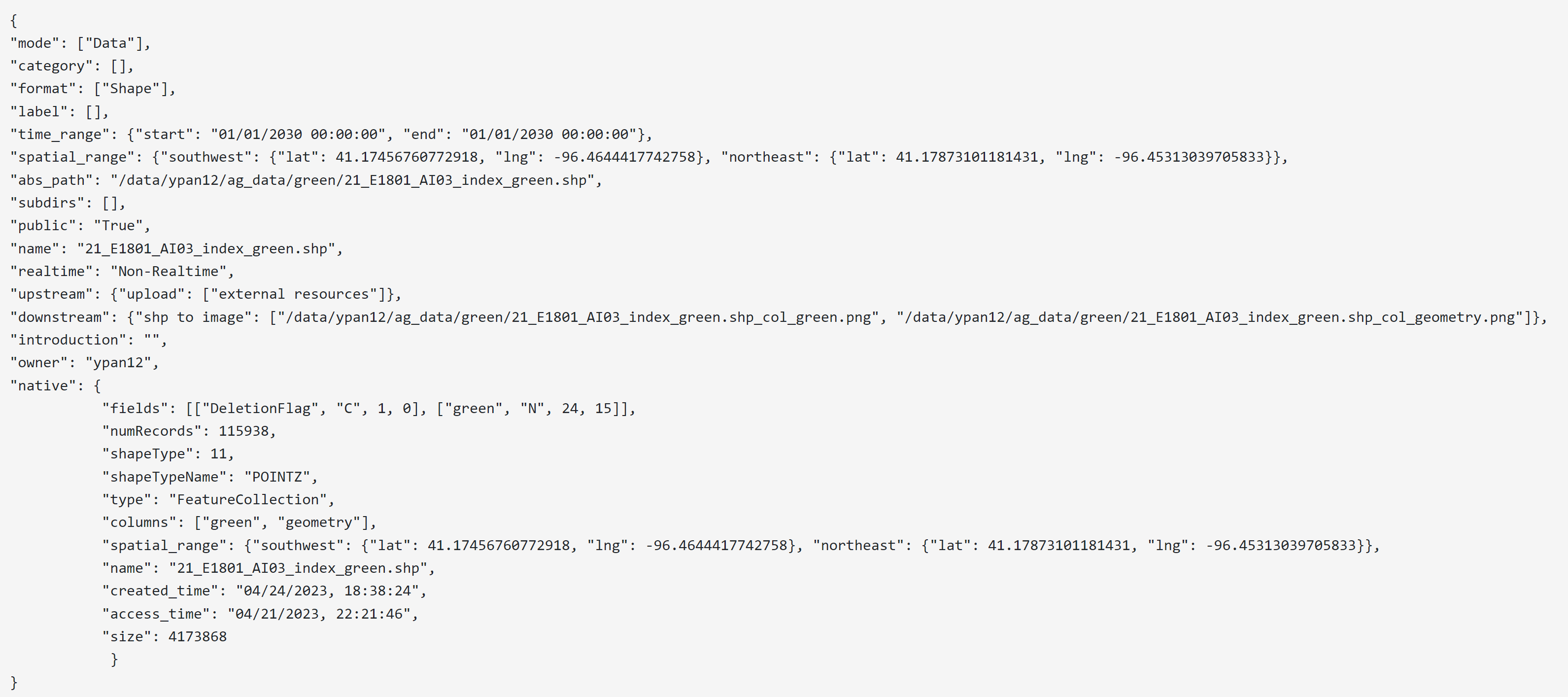}
        \caption{Retrieving Meta Data by Path}
        \label{fig:api_sub1}
    \end{subfigure}%
    \hfill
    \begin{subfigure}{0.4\textwidth}
        \centering
        \includegraphics[width=\linewidth]{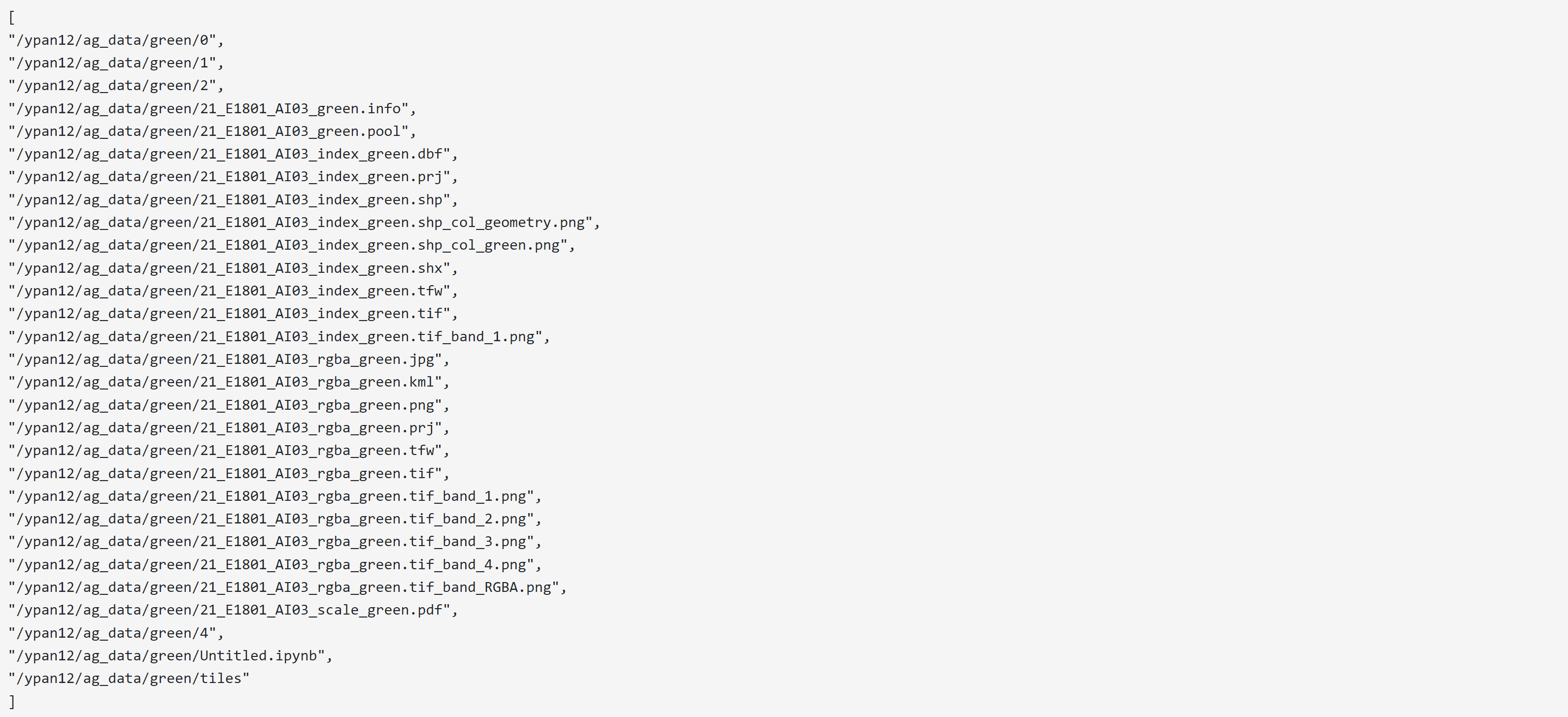}
        \caption{Retrieving Directory Structure by Path}
        \label{fig:api_sub2}
    \end{subfigure}%
    \hfill

    \caption{JSON Files Returned by APIs}
    \label{fig:api}
\end{figure}

ADMA exposes several data APIs to facilitate data retrieval by authorized third-party programs. Data APIs are in REST style, and returned values are in human-readable JSON format. Using URLs in the format \texttt{/api\_meta\_data?path=\{path\}\&key=\{key\}}, a user can retrieve the metadata by path. For example, if the request URL contains:

\noindent \texttt{\scriptsize{path=/username/ag\_data/green/21\_E1801\_AI03\_index\_green.shp}}

\noindent and given a valid key, we can retrieve the metadata for the specific file \texttt{21\_E1801\_AI03\_index\_green.shp}.
The returned metadata will be in JSON format as shown in Figure \ref{fig:api_sub1}. 
Using URLs with the API \texttt{/api\_list\_sub\_items?path=\{path\}\&key=\{key\}} will list files or directories within any \texttt{\{path\}} which are visible to the user when the a \texttt{\{key\}} is specified. For example, if the request URL contains \texttt{path=/username/ag\_data/green}, and a valid key is given, we can retrieve all the files and folders inside the directory: \texttt{/username/ag\_data/green}. The returned metadata will be in JSON format as illustrated in Figure \ref{fig:api_sub2}.

\subsubsection{Data Privacy and Authentication}

ADMA protects data privacy by differentiating private data from public data. Users can choose to publish their data by toggling the locker icon and updating the metadata of a folder or file. Once a file is made public, all its sub-folders and files within it will be recursively made public, and vice versa. All the public data can be viewed under the folder: \texttt{/username/ag\_data/public\_data}. Each user is assigned a separate root directory for their data, which is further consolidated by a containerized environment for running user-defined tools and Jupyter Notebook.

\subsection{Comparison}

By adopting similar evaluation frameworks as in \cite{runck2022digital}, we compare ADMA with some of the existing agriculture platforms. The dimensions of evaluation criteria and corresponding explanation are listed in Table \ref{tab:criteria}. The chosen criteria cover a broad range of dimensions, from the basic requirements for a typical data management platform to more innovative features, which also take into consideration the principles of FAIR. The ten dimensions of criteria are, namely, privacy and security, CRUD repository, interoperability, extensibility, data model, data granularity, analytical tools, trackability, ML Models, and Intelligence.

By using the listed criteria, we make a comparison with existing public agriculture data platforms such as CyVerse, GARDIAN, GEMS, and TERRAREF. The comparison results are listed in Table \ref{tab:comparison}. We mark $\times$ when a platform satisfies certain criteria and leave the cell empty otherwise. We can see all of the platforms satisfy the criteria such as privacy and security, CRUD repository, interoperability, extensibility, and analytical tools. Some of the platforms support the criteria of data model, data granularity, and ML models, and only our platform supports trackability and intelligence. Our model supports all the considered criteria. Although this is merely a qualitative comparison and bias may be introduced due to the choice and the specific definition of evaluation criteria, the comparison results can provide some clues that our platform has superiority in the chosen dimensions of criteria and thus the designing ideology of our system cover a more vast area of potential requirements and interest for the users in the future.   

\begin{table*}[htbp]
\caption{Dimensions of Evaluation Criteria}
\begin{center}
\begin{tabularx}{\linewidth}{|>{\hsize=.3\hsize}X|>{\hsize=1.7\hsize}X|}
\hline
\textbf{Dimension}&{\textbf{Explanation}} \\
\hline
Privacy\&Security  & The system protects data privacy and security, including but not limited to: 1. the ability to separate data and metadata; 2. the ability to share data or tools with other users safely; 3. the ability to differentiate public and private data; 4. the ability to run user-defined programs safely; 5. unified authentication module; 6. data encryption.\\
\hline
CRUD Repository & Data can be created (uploaded), read (rendered), updated, and deleted from/to/on the repository.  \\
\hline
Interoperability&  Data from external sources of different formats or various modalities can be added or incorporated into the system and managed by the system without difficulties.  \\
\hline
Extensibility &  New capabilities and functionalities can be added to the system. \\
\hline
Data Model &  Extract, transform and load procedure employed to populate a relational, graph, object, or vector database governed by an explicit data model. \\
\hline
Data Granularity & Ability to index and query for sub-file level entities such as individual records within datasets.\\
\hline
Analytical Tools &  Provides tools for users to clean text, numeric, and geospatial outliers, as well as conduct unique analysis.\\
\hline
Trackability & Ability to keep track of different operations on each file, maintain and visualize the pipeline record accordingly.\\
\hline
ML Models & Provide ready-to-use or support training and hosting of agricultural models.\\
\hline
Intelligence & The system utilizes Artificial Intelligence to facilitate data management, analysis, and interaction.\\
\hline
\end{tabularx}
\label{tab:criteria}
\end{center}
\end{table*}

\begin{table*}[t]
\caption{Comparison of Agriculture Data Management Platforms across a Variety of Criteria}
\begin{center}
\scalebox{0.75}{
\begin{tabular}{|c|c|c|c|c|c|c|c|c|c|c|}
\hline
\textbf{Platform} & \textbf{Privacy\& Security}& \textbf{CRUD Repository}& \textbf{Interoperability} & \textbf{Extensibility} & \textbf{Data Model} &\textbf{Data Granularity} & \textbf{Analytical Tools}& \textbf{Trackability} &\textbf{ML Models}& \textbf{Intelligence}\\
\hline
Cyverse & $\times$ & $\times$ & $\times$ & $\times$ &  & $\times$ & $\times$ &  & $\times$ & \\
\hline
GARDIAN & $\times$ & $\times$ & $\times$ & $\times$ &  &  & $\times$ &  &  & \\
\hline
GEMS & $\times$ & $\times$ & $\times$ & $\times$ &  &  & $\times$ &  & &\\
\hline
TERRAREF & $\times$ & $\times$ & $\times$ & $\times$ & $\times$ & $\times$ & $\times$ &  & & \\
\hline
ADMA & $\times$ & $\times$ & $\times$ & $\times$ & $\times$ & $\times$ & $\times$ & $\times$ & $\times$ & $\times$\\
\hline

\end{tabular}
}
\label{tab:comparison}
\end{center}
\end{table*}
\section{Conclusion}
\label{sec:conclusion}

In this paper, we delved into the multifaceted challenges that modern agriculture confronts, particularly in the context of burgeoning global demands, climate change, and the constraints of dwindling resources. Central to addressing these challenges is the effective management and analysis of vast, heterogeneous data, which has become increasingly complex with the proliferation of sensors, IoT devices, and advanced instrumentation. Recognizing the inefficiencies and limitations of existing data management practices, we have introduced Agriculture Data Management and Analytics (ADMA), a pioneering platform tailored for agricultural domains. ADMA, grounded in the FAIR principles, offers a holistic solution that amalgamates interactivity, scalability, flexibility, and open-source capabilities, ensuring that data from diverse agricultural disciplines is not only accessible but also actionable. By bridging the gap between data collection and actionable insights, ADMA stands as a testament to the transformative power of data-driven decision-making in agriculture. As we look ahead, it is evident that platforms like ADMA will be instrumental in shaping a sustainable, resilient, and productive future for agriculture, ensuring that we can meet the world's growing demands while navigating the complexities of our changing environment.

\section*{Acknowledgment}
 This research is funded by the Nebraska Research Initiative (NRI).

\bibliographystyle{IEEEtran.bst}
\bibliography{IEEEabrv}

\end{document}